\documentclass[12pt]{iopart}
\pdfoutput=1

\usepackage{algpseudocode}
\usepackage{pgfplots}
\usepackage{adjustbox}
\expandafter\let\csname equation*\endcsname\relax
\expandafter\let\csname endequation*\endcsname\relax
\usepackage{amsmath, amsthm, amssymb, amsfonts}
\usepackage{multirow}
\usepackage{footnote}
\usepackage[nocompress]{cite}
\usepackage{tikz}
\usetikzlibrary{positioning}
\usepackage{url}

\usepackage{float}
\usepackage{subfig}
\usepackage{nameref,hyperref}
\usepgfplotslibrary{groupplots}

\newtheorem{theorem}{Theorem}

\newtheorem{definition}{Definition}

\usepackage[shortlabels]{enumitem} %enumerate abc
\newcommand{\rbrac}[1]{\left( #1 \right)}

\newcommand{\sbrac}[1]{\left[ #1 \right]}
\newcommand{\cbrac}[1]{\left\{ #1 \right\}}
\newcommand{\eps}{\epsilon}
\newcommand{\diag}{\text{diag}}
\renewcommand{\a}{\alpha}

\renewcommand{\a}{\alpha}

\renewcommand{\a}{\alpha}

\newcommand{\DDb}[1]{{\em \textcolor{blue}{#1}}}

\renewcommand{\P}{\mathbb{P}}

\makeatletter

\def\env@sqcases{%
	\let\@ifnextchar\new@ifnextchar
	\left\lbrack
	\def\arraystretch{1.2}%
	\array{@{}l@{\quad}l@{}}%
}
\makeatother

\begin{document}

\title[A principled framework of functional connectome thresholding]{A principled framework to assess the information-theoretic fitness of brain functional sub-circuits}
\author{
    Duy Duong-Tran$^{1,2,*,\ddagger}$, 
    Nghi Nguyen$^{3, \ddagger}$, 
    Shizhuo Mu$^{1}$, 
    Jiong Chen$^{1}$, 
    Jingxuan Bao$^{1}$, 
    Frederick Xu$^{1}$, 
    Sumita Garai$^{1}$, 
    Jose Cadena-Pico$^{4}$, 
    Alan David Kaplan$^{5}$, 
    Tianlong Chen$^{6}$,
    Yize Zhao$^{7}$,
    Li Shen$^{1,\dagger}$, 
    and Joaqu\'{i}n Go\~{n}i$^{8,9,10,\dagger}$
}

\address{
$^{1}$ \quad Department of Biostatistics, Epidemiology, and Informatics (DBEI), Perelman School of Medicine, University of Pennsylvania, Philadelphia, PA, USA\\ 
    $^{2}$ \quad Department of Mathematics, United States Naval Academy, Annapolis, MD, USA\\
    $^{3}$ \quad Gonda Multidisciplinary Brain Research Center, Bar-Ilan University, Ramat Gan, Israel\\
    $^{4}$ \quad Machine Learning Group, Lawrence Livermore National Laboratory, Livermore, CA, USA\\
    $^{5}$ \quad Computational Engineering Division, Lawrence Livermore National Laboratory, Livermore, CA, USA\\
    $^{6}$ \quad Department of Computer Science, The University of North Carolina at Chapel Hill \\
    $^{7}$ \quad School of Public Health, Yale University, New Heaven, CT, USA   \\
    $^{8}$ \quad School of Industrial Engineering, Purdue University, West Lafayette, IN, USA\\
    $^{9}$ \quad Purdue Institute for Integrative Neuroscience, Purdue University, West Lafayette, IN, USA\\ 
    $^{10}$ \quad Weldon School of Biomedical Engineering, Purdue University, West Lafayette, IN, USA \\
* Correspondence to: duongtra@usna.edu \\
$\ddagger$ Equal contribution \\
$\dagger$ Co-supervising Authors}
% \ead{duongtra@usna.edu}
\vspace{10pt}
%\begin{indented}
%\item[]August 2017
%\end{indented}

\begin{abstract}
    In systems and network neuroscience, many common practices in brain connectomic analysis are often not properly scrutinized. One such practice is mapping a predetermined set of sub-circuits, like functional networks (FNs), onto subjects' functional connectomes (FCs) without adequately assessing the information-theoretic appropriateness of the partition. Another practice that goes unchallenged is thresholding weighted FCs to remove spurious connections without justifying the chosen threshold. This paper leverages recent theoretical advances in Stochastic Block Models (SBMs) to formally define and quantify the information-theoretic fitness (e.g., prominence) of a predetermined set of FNs when mapped to individual FCs under different fMRI task conditions. Our framework allows for evaluating any combination of FC granularity, FN partition, and thresholding strategy, thereby optimizing these choices to preserve important topological features of the human brain connectomes. 
   By applying to the Human Connectome Project with Schaefer parcellations at multiple levels of granularity, the framework showed that the common thresholding value of 0.25 was indeed information-theoretically valid for group-average FCs despite its previous lack of justification.
    Our results pave the way for the proper use of FNs and thresholding methods and provide insights for future research in individualized parcellations.
\end{abstract}

\section{Introduction}

The success of large-scale brain connectomics---which subserves a myriad of neuroimaging research endeavors based on fMRI \cite{tang2018connectome,mijalkov2020delayed,zhan2017significance}, MEG \cite{colclough2015symmetric}, and EEG \cite{thilaga2015heuristic}---hinges on choosing representations of functional connectivity that are as well-defined as possible. Functional connectomes (FCs) are often constructed by computing a statistical dependency measure, such as the Pearson correlation coefficient, across all specified pairs of the brain's regions of interest (ROIs) using the aggregated voxel level blood-oxygen-level-dependent (BOLD) signals. However, constructing FCs from BOLD signals with activation delays (due to inhibitory-excitatory dynamics possibly causing negative ROI correlations) can significantly impact estimates of population-level FCs \cite{tang2018connectome} and the associated functional brain network topological features such as nodes' centrality \cite{alakorkko2017effects}, global network measures \cite{mijalkov2020delayed}, and geometry-topology relation \cite{roberts2016contribution}. Recent efforts have focused on improving FC construction by taking into account neuronal signal activation delays \cite{mijalkov2020delayed} and negative correlations \cite{zhan2017significance}. Nonetheless, much effort is still needed to quantify the efficacy of each FC construction framework, especially in terms of preserving the ``true'' features of the population FCs that shed light on fundamental principles of the brain.

Functional sub-circuits, \textit{e.g.}, functional networks (FNs) \cite{yeo2011organization}, and their modularity characteristics \cite{meunier2010modular,stam2010emergence,bertolero2015modular,duong2019morphospace} are crucial to understanding such fundamental neural principles, including brain complexity \cite{bassett2011understanding}, differential configurational properties \cite{duong2019morphospace}, modular structures \cite{sporns2016modular,meunier2010modular}, and information processing \cite{amico2019centralized,amico2019towards}. Studies on the modular organizations of the human brain have also informed applied research on aging \cite{meunier2009age,betzel2014changes} and disorders including schizophrenia \cite{alexander2012discovery}. Moreover, research consistently shows that executive subsystems in the brain are reproducible across many individuals at rest, \textit{e.g.},  \cite{power2011functional,yeo2011organization}, indicating a widespread application of these FNs in various studies—from control groups \cite{lee2012clustering} to pathological investigations \cite{chan2017resting} and predicting individual differences \cite{reineberg2015resting}.  Even so, there have been few (if any) systematic studies addressing the validity of a common and rarely challenged practice in brain connectomics, which is applying one specific set of \textit{a priori} FNs to multiple FCs. In other words, FC processing usually involves mapping an \textit{a priori} fixed set of FNs onto the constructed FCs, across different subjects and fMRI task conditions, without examining whether those mappings are relevant information-theoretically fit to the constructed FCs, given the existence of human brain fingerprint \cite{abbas2020geff,abbas2023tangent,duong2019morphospace,chiem2022improving,finn2015functional,amico2018quest,abbas2020geff,garai2023mining,garai2024quantifying,amico2021toward,amico2019towards}.

Among the many decisions influencing whole-brain functional connectivity estimates like FCs and circuit-level representations like FNs, the choice of brain parcellations, \textit{i.e.}, how nodes in functional brain networks are defined, is undoubtedly one of the most critical steps. \cite{schaefer2018local,glasser2016multi,salehi2018there}. In fact, this choice determines the network topology used in downstream analyses. Recent studies have shown that different levels of parcellation granularity can affect the identification of subject-level FC fingerprints \cite{finn2015functional,abbas2020regularization}. In an effort to register the raw neuroimaging data into a sequence of increasing granularity, Schaefer and colleagues have recently published a scheme of atlases that increase in network sizes. These parcellations refine the robust set of resting state networks initially identified by Yeo et al. \cite{yeo2011organization}, offering various granularity levels for in-depth analysis. Thanks to these advancements, the brain connectivity research community can now explore characteristics of sequential functional brain networks, especially those coupled with the corresponding \textit{a priori} set of FNs.

Regardless of which parcellation scheme is employed during large-scale FC and FN analyses, another common practice in network neuroscience is thresholding (or, more generally, eliminating statistically spurious functional edges) based on some arbitrary rules or research hypotheses. Careful design of the thresholding process is central to ensuring scientific rigor not only in healthy control studies but also in those studying disorders such as schizophrenia \cite{vavsa2018probabilistic}, unipolar depression, and bipolar disorder \cite{yu2020abnormal}. Unrigorous application of thresholding can therefore undermine the validity of such important studies by affecting downstream analyses, including parametric statistical tests \cite{langer2013problem} and network characterization \cite{van2017proportional}. To mitigate such issues, various thresholding strategies have been proposed to retain particular desired attributes of the original weighted networks. These strategies include proportional thresholding aimed at keeping the absolute number of edges across different subjects and tasks \cite{van2017proportional}, modular similarity \cite{yu2020abnormal}, and percolation aimed at preserving the topological features of the original weighted graph \cite{esfahlani2018percolation}. Spurious edge elimination also involves methods based on wavelets \cite{vavsa2018probabilistic}, mixture modeling \cite{bielczyk2018thresholding}, topological data analysis through persistent homology \cite{lee2011discriminative,lee2012persistent}, branch-and-bound based algorithms (to study cognitive activity \cite{thilaga2015heuristic}), and orthogonal minimal spanning trees for dynamical functional brain networks \cite{dimitriadis2017topological}. Furthermore, alternatives to thresholding treatment for FCs have also been proposed using hierarchical Bayesian mixture models. \cite{gorbach2020hierarchical}. However, this multitude of strategies further complicates the already complex decision-making process of brain data preprocessing and analysis. After all, how can one determine which combination of FC parcellation, FN partitioning, and edge pruning techniques is optimal for their dataset? {To the best of our knowledge, no studies have offered a mathematically justified and robust process for choosing such combinations.

This work tackles the complexity posed by that abundance of choices and lack of theoretical justification}. Our objectives are two-fold: \textit{i)} formalizing and quantify the level of information prominence of a given fixed set of FNs across different subjects and tasks, and \textit{ii)} using the level of prominence as guidance to eliminate spurious functional edges in whole-brain FCs. To do so, we utilize Schaefer parcellations \cite{schaefer2018local} with nine distinct granularity levels, ranging from 100 to 900 nodes in 100-node increments. We first present an overview of Schaefer parcellations, as well as a formalization of stochastic block models (SBMs) and its relevance to our quest \textbf{Section 2}. We then propose an SBM reconstruction pipeline in \textbf{Section 3}. We wrap up with Results (\textbf{Section 4}) and Discussion (\textbf{Section 5}). Our framework can be generalized to any given pair of an FN partition and a parcellation (\textit{e.g.} \cite{glasser2013minimal,tian2020hierarchical}).

\section{A principled framework to assess information theoretical fitness of brain functional sub-circuits/networks}

\subsection{{Schaefer's atlas of cerebral cortex}}
The brain parcellations used in this work are sequential, in the sense that their granularity increases, ranging from 100 nodes to 900 nodes with increments of 100 nodes, and are registered on the cortical surface of the brain. These sequential atlases are made possible thanks to the work of Schaefer and colleagues \cite{schaefer2018local}. To ensure our framework is consistent with prior works in human brain connectomics \cite{amico2018quest,amico2018mapping}, we added 14 sub-cortical regions, resulting in network sizes of 114, 214, ..., 914 nodes for all fMRI conditions in the Human Connectome Project (HCP) repository, Released Q3 \cite{van2012human,van2013wu}. Schaefer parcellations are subdivisions of Yeo's functional networks \cite{yeo2011organization}, in such a way that any coarser-grained partition is a subdivision of or related to a finer-grained Schaefer one. For instance, let $u_{114}$ be a node in the Schaefer graph with $n=114$ nodes. Every node is further subdivided into two nodes $v'_{214}$ and $v''_{214}$ in the next Schaefer graph in the sequence, i.e., the one with $n=214$ nodes.

In practice, the subsequent divisions from coarser to finer granularity of Schaefer parcellations are not perfectly hierarchical, in the sense that one brain region in the coarser parcellation is not perfectly parcellated into subsequently smaller regions the finer one. Nonetheless, for relaxation purposes, if a node associated with the coarser parcellation has significant spatial overlaps with those in subsequently finer parcellations of the Schaefer graph sequence, they are assigned to the same resting state network.

\subsection{Stochastic Block Models (SBMs)}
Stochastic Block Models (SBMs) have recently gained traction due to exciting developments in both theoretical and practical domains (see Preliminaries-Stochastic Block Models in Supplementary Information for further details on notations and a brief introduction). Theory-wise, phase transitions in the fundamental limits of community \textit{detection} (or more generally, mesoscopic structures) were discovered through the measure Signal-to-Noise Ratio (SNR) \cite{abbe2017community}. In the domain of brain connectivity, SBM has demonstrated its advantages in exploring and uncovering diverse types of brain functional sub-circuits (\textit{e.g.}, dis-assortative or core-periphery) beyond the traditional assortative mesoscopic structures \cite{betzel2018diversity,faskowitz2018weighted}. Specifically, Sandon and Abbe, in \cite{abbe2017community}, laid out a comprehensive treatment of criteria for mesoscopic structure recovery for any pair of a networked system and an \textit{a priori} set of communities (or functional networks in brain connectomic domain). Specifically, the recovery requirements were classified under: 
\begin{enumerate}
	\item Weak Recovery (also known as community detection); 
	\item Almost Exact Recovery;
	\item Exact Recovery.
\end{enumerate}

The recoverability of the ground-truth partition depends on the degree regime (indicated by the degree scaling factor $s_t$) in which the network resides. For instance, weak recovery only requires the necessary condition for a limiting graph ($n\rightarrow \infty$) to be in the constant degree regime, \textit{i.e.}, $O(\frac{1}{n})$. On the other hand, exact recovery requires the necessary condition (for limiting graph) that the graph is asymptotically connected, \textit{i.e.}, in the degree regime of logarithmic $O(\frac{log(n)}{n})$. The sufficient condition for all the recovery criteria is stated in the respective theorems with different proposed measures with sharp phase transitions, as seen in \cite{abbe2017community}.\footnote{If a measure (say for weak or exact recovery) is below a certain algebraic threshold (stated in the respective theorems), recovery is not possible although the necessary condition is satisfied.} Further details on recovery theorems are located in Supplementary Information.         

Here, we chose the weak-recovery requirements as guidance for whole-brain functional connectivity estimation for four reasons:
\begin{enumerate}
	\item Although Schaefer parcellations with an increasing number of nodes allow us to project some empirical insights onto their degree regime, a rigorous theoretical argument on the degree regime is not possible for any empirical graph sequence. Hence, exact recovery of an \textit{a priori} unique ground-truth partition is not relevant in the case of brain functional connectomes;
	\item Even in the empirical domain, we observe that both group-average and individual FCs become disconnected (\textit{i.e.}, the number of connected components is more than 1) after a relatively small threshold value in the interval $\tau \in [0.2,0.3]$. Theoretically, a graph sequence is required to be connected, \textit{asymptotically}, to fulfill the requirements for \textit{exact recovery}. On the other hand, weak-recovery (detection of mesoscopic structures) offers a more realistic and relaxed set of criteria for this particular application. This facilitates estimating a whole-brain FC that is most suitable for an \textit{a priori} set of FNs without evaluating the number of connected components of the thresholded FC.  
	\item Most (if not all) mesoscopic studies of brain functional sub-circuits such as \cite{betzel2018diversity,faskowitz2018weighted} are based on pre-defined hypotheses, \textit{e.g.}, that the brain functional sub-circuits involve a more diverse class of community than just assortative ones \cite{betzel2018diversity}. Such assumption leads to the appropriate usage of different community detection algorithms such as Weighted Stochastic Block Models (WSBM) in the case of \cite{betzel2018diversity,faskowitz2018weighted}. As mentioned above, weak-recovery is equivalent to community detection in the theoretical SBM literature;
	\item No set of functional sub-circuits is universally agreed and uniquely identified as the \textbf{ground-truth communities}. Hence, all proposed brain functional sub-circuit parcellations,\textit{e.g.}, \cite{yeo2011organization}, are relative.
\end{enumerate}
	\subsection{SBM description, inference and extended usage}
\subsubsection{Model Description}
In this section, we define some of the key components of SBM. Other fundamental mathematical notations are referred to in the section Stochastic Block Model Preliminaries in Supplementary Information.
\begin{itemize}
	\item $G=[a_{uv}]=\begin{cases} FC,& \text{weighted-graphs}\\ M,&\text{binarized-graphs}\end{cases}$: network/graph (\textit{e.g.} FCs in the context of this work);
	% \item $G_\t$ is the thresholded network where $$G_\t=\begin{cases} 0,& |a_{uv}|<\t\\ \\\begin{sqcases} 1,& |a_{uv}|\ge \t,\forall M_t\\ |a_{uv}|,& |a_{uv}|\ge \t,\forall FC_\t \end{sqcases} \end{cases}$$
	\item $V(G)=\cbrac{u}$, and $E(G)=\cbrac{uv\mid u,v\in V(G)}$ be set of vertices and edges, respectively;
	\item The size and order of a network are denoted by $|V(G)|=n$ and $|E(G)|$, respectively;
	\item $\cbrac{G_t}, \forall t\in \mathbb{N}$ is the graph sequence; in the empirical domain, the number of graphs in the sequence is defined as $|\cbrac{G_t}|=T$; 
 %\DDb{n -> not correct - use a different symbol for sequence index} \NNm{n changed to t, N changed to T}
	\item $k$ is the number of communities/clusters;
	\item $\sigma=[\sigma_u]\in [k]^n$ is the pre-defined, well-understood community assignment in vector form of length $n$. It is the mathematical map $\sigma = \cbrac{ u \mapsto i,\forall u\in[n],i\in [k]}$. In general, $\sigma$ is also referred to as a graph partition;	
	\item $\Omega=[|\Omega_i|]$ is the vector containing cardinality of community where $$|\Omega_i|=|\cbrac{u\mid \sigma_u=i}|,\forall i\in [k],u\in[n]$$
	\item $C$ is the statistical summary of edge properties within and between communities in matrix form. Mathematically, 
	$$C=	\begin{cases}
	C_{bin} \in \mathbb{N}^{k\times k}_{+}\\
	C_{wei} \in \mathbb{R}^{k\times k}
	\end{cases}$$
	where $C_{bin} \in \mathbb{N}^{k\times k}_{+}$ denoted the simple edge count matrix within or between communities and $C_{wei}$ denoted the weighted edge sum (also within or between communities);
	\item $C_{max}\in \mathbb{N}^{k\times k}_{+}$ is the maximum number of edges within or between communities;
	\item $p=[p_i]$: the probability that a node $u$ belongs to community $i\in [k]$; $P=\diag(p)$ is a $k\times k$ matrix filled with $p_i$ in the diagonal;
	\item $Q=[Q_{ij}]\in \mathbb{R}^{k\times k}$ is the expected node degree matrix, \textit{i.e.} the expected number of connections a node in community $i$ has with community $j$;
	\item $s_t$: scalable factor of degree regime in a graph sequence ${G_t}$ where $t\in T$;
	\item $W_{s_t}=[w_{ij}]_{s_t}$ is the edge probability between 2 nodes in community $i$ and $j$ in terms of the scaling factor $s_t$.\footnote{It is worth noting
 that if $w_{ij}$ is the same for all $i,j\in [k]$, then SBM collapses to classical ER random graph model} We use $W$ to denote the edge probability matrix with $s_t = 1$;
	\item $PQ=nP\sbrac{\frac{W}{s_t}}=nPW$ is the community profile matrix where $i$ column is the expected number of edges that community $i$ has with all communities. Note that for weak-recovery (detection), scaling factor $s_t=1$.
\end{itemize}

\subsubsection{Inference and extended usage}
The basis of SBM parameter inference is reverse engineering by the maximum likelihood principle. Specifically, since both $G$ and $\sigma$ (subsequently, $k=\max_{u\in [n]} \sigma_u$) are priors, in expectation, we can infer $SBM(P,W)$ using the Bayesian approach as follows: 
\begin{enumerate}
	\item $P=\frac{\Omega}{n}=[p_i]=\sbrac{\frac{|\Omega_i|}{n}}$
	\item Infer $W_{bin}=\frac{C_{bin}}{C_{max}}$ 
	\item Compute $W_{wei}=\frac{C_{wei}}{C_{max}}$
	\item $Q=nW$ as $s_t=1$ for weak-recovery
	\item Compute $PQ$ (\textit{Matrix Multiplication})
\end{enumerate}
where $C_{bin}$ is a simple edge count of $M^{GA}_\tau$ between or within blocks of communities whereas $C^{wei}$ is the sum of weighted edges of $FC_\tau$ (also between or within communities). Specifically,
$$
C_{bin} = \sum_{u,v\in[n]} \mathbf{1}_{\sigma_u=\sigma_v}
$$
$$
C_{wei} = \sum_{u,v\in[n]} |\mathbf{w}_{uv}|, \sigma_u = \sigma_v
$$
and 
$$
C_{max} = \Omega \Omega^T
$$
The inference of matrix $P$ is based on the law of large numbers \cite{abbe2017community}. For $W_{bin}$, we perform entry-wise divisions of matrix $C_{bin}$ by matrix $C_{max}$, which infers the Bernoulli random variable parameter $p$ representing the probability of successful edge formation between each pair of stochastically equivalent nodes within or between communities. In the case of $C_{wei}$, note that we use the term \textbf{computing} instead of inferring because we have extended the usage of $SNR$ to mesoscopic prominence measure. We use the absolute values $\|\mathbf{w}_{uv}\|$ to only consider the overall magnitude (and not the sign) of functional couplings within/between FNs. 

Technically, this inference is less challenging than traditional inference problems where $\sigma$ is also a latent variable in the model and graph ensemble $G$ is the only observable ensemble available. Specifically,
$$
(G,n,\sigma,k) \sim SBM(P,W)
$$
where $G$ and $\sigma$ are priors.
	\subsection{Weak Recovery of ground-truth partition}
\begin{definition}
	Weak recovery of a ground-truth partition can be rigorously equivalent to the existence of an algorithm that infers a partition that agrees with the ground-truth one up to $\max_{i} p_i + \eps, \forall i\in [k]$. This level of accuracy is the minimal requirement for most community detection methods.
\end{definition}
\begin{theorem}
	(Sandon and Abbe \cite{abbe2017community}) Let $(G,\sigma) \sim SBM\rbrac{n,p,\frac{s_t Q}{n}}$ for $p,Q$ arbitrary and $s_t=1$. If $SNR > 1$, then weak recovery is efficiently solvable; where
	$$
	SNR = \frac{\lambda_2^2}{\lambda_1}
	$$
	and $\lambda_i$ is the $i^{th}$ eigen value of the community profile matrix $PQ$.	
\end{theorem}

Weak recovery of given ground-truth communities means that through that algorithm, the recovered partition outperforms a random guess, \textit{i.e.} $\max_{i} p_i$, by a small factor $\eps$. The criteria for weak recovery are driven by a hard threshold approach presented in the below theorem. Importantly, achieving weak recovery does not necessitate the graph being connected under an asymptotic regime. Loosely speaking, we only need every graph in the graph sequence to have a large connected component. In other words, we only need $\cbrac{G_{t\in T}}$ to be in the constant degree regime, \textit{i.e.} $s_t=1$. Consider a network of $n$ nodes divided into two equal-sized ground-truth communities (\textit{i.e.} $\frac{n}{2}$ nodes for each community). In a weak recovery scenario, it is feasible to accurately identify the community membership of each node with a probability marginally above $50\%$, say by an additional $5\%$. It implies that if an ensemble is generated under a constant degree regime, one can arbitrarily assign any community membership to isolated nodes, \textit{i.e.} leaves; hence, exact recovery is impossible in this regime. On the other hand, for exact recovery, since $W$ scales with $n$ through the factor $s_t$, the community profile matrix $M$ consequently grows with the factor $s_t$ as well.
% Begin pseudo-code
\begin{figure*}[!ht]
	\footnotesize
	\begin{algorithmic}[1]
		\Procedure{reconFC} {$\Gamma$, $n$, $k$, $\sigma$, $\mathbb{S}$, $P$, $W$ \emph{FC,M}}
    		\For{every Schaefer Granularity level and fMRI task}
        		% \State $\rightarrow$ \textbf{Vetting Step}: Step 1 and 2
        		\State \DDb{\textbf{Step 1: Compute the group-average FC ($FC^{GA}$)}} 
                    \State Using all individual FCs ($FC^\gamma$) given a Schaefer parcellation and fMRI task, 
        		$$
        		FC^{GA}=\frac{\sum_{\gamma=1}^\Gamma FC^\gamma}{\Gamma}
        		$$
                    \State \DDb{\textbf{Step 2: Vetting}}
        		\For{$\vec{\tau}\in \mathbb{S}$}
        			\State \emph{Compute} the masked, binarized group-average FC: $M^{GA}_{\vec{\tau}}=
                        \begin{cases}
            			1, & |FC^{GA}|\succ \vec{\tau} (*)\\
            			0, &   o.w.
        			\end{cases}$
        			\State Infer SBM parameters and apply \textbf{Theorem 1} to compute $SNR[M^{GA}_{\vec{\tau}}]$
                    \EndFor
        		\State Determine the weak-recoverability sub-interval $I =\displaystyle \prod\limits_{i=1}^{\text{dim}(\mathbb{S})} [a_w^{(i)}, b_w^{(i)}]$ by 
        		$$a_w =\underset{{\vec{\tau}} \in \mathbb{S}}{\text{argInf}} (SNR[M^{GA}_{\vec{\tau}}]>1)$$
        		$$b_w =\underset{{\vec{\tau}} \in \mathbb{S}}{\text{argSup}}(SNR[M^{GA}_{\vec{\tau}}]<1)$$
                    \State \DDb{\textbf{Step 3: Compute the \emph{mesoscopic prominence measure} for each individual FC}}
        		\For{$\gamma\in [\Gamma]$}
        			\For{${\vec{\tau}} \in \mathbb{S}$}
            			\State Compute the individual thresholded weighted FC: $FC^\gamma_{{\vec{\tau}}}=
                            \begin{cases}
                			|FC^{\gamma}|, & |FC^{\gamma}| \succ \vec{\tau} (*)\\
                			0 &  o.w.
            			\end{cases}$ 
            			\State Compute the \emph{mesoscopic prominence measure} $SNR[FC^\gamma_{\vec{\tau}}]$ using \textbf{Theorem 1}.
        			\EndFor
                    \EndFor
                    \State \DDb {\textbf{Step 4: Obtain optimal thresholding parameters and check their weak-recoverability}}
                    \For{$\gamma\in [\Gamma]$}
            		\State Obtain $\vec{\tau}_{opt}=\underset{\vec{\tau} \in \mathbb{S}}{\text{argmax}}(SNR[FC^\gamma_{\vec{\tau}}])$
            		\State Check if $\vec{\tau}_{opt}\in I =\displaystyle \prod\limits_{i=1}^{\text{dim}(\mathbb{S})} [a_w^{(i)}, b_w^{(i)}]$
                    \EndFor
    		\EndFor
		\EndProcedure
	\end{algorithmic}
	\caption{Pseudo-code for \emph{reconFC} routine using the number of individual FCs $\Gamma$, Schaefer granularity $n$, number of functional networks $k$, \textit{a priori} partition $\sigma$, parameter space $\mathbb{S}$, community assignment likelihood $P$, and connectivity pattern matrix $W$. {Note that $\mathbb{S}$ could contain $dim(\mathbb{S})$ parameters in general. In addition, in line 14, the notation $G \succ \vec{\tau} (*)$ represents a particular configuration of matrix $G$ that satisfies the parameter space $\mathbb{S}$ represented by the value of $\vec{\tau}(*)$.}
 }
	\label{alg:NGSC}
\end{figure*}

\subsection{Problem Formulation and the fitness assessment framework}
\subsubsection{Problem formulation:}
\noindent {\textbf{Problem Statement:} Given an \textit{a priori} set of functional sub-circuits, we would like to investigate and subsequently quantify the "fitness" of such pre-determined partition for a given brain network topology.}

\noindent {\textbf{Problem Formulation:} We formulate this problem as the "dual" of the primal community detection problems. \cite{fornito2015connectomics,fortunato2010community,fortunato2016community,fornito2016fundamentals}. Specifically, in the primal problem, a partition (\textit{i.e.,} set of mesoscopic structures) is constructed for a given network topology. In the dual, we look to evaluate and subsequently quantify the fitness of a pre-determined (\textit{e.g., a priori}) set of mesoscopic structures for the given network. }

\subsubsection{Fitness Assessment Framework:}
For a given pair of a complex network (\textit{e.g.}, functional connectome) and an \textit{a priori} set of ground-truth communities (\textit{e.g.}, Yeo's functional sub-circuits), we propose the following steps to access the information-theoretic fitness of ground-truth communities (Figure \ref{alg:NGSC}) as follows:
\begin{enumerate}
    \item Step 1: Obtain an average representation (\textit{e.g.}, a group-average FC) from the collection of individual networks, binarize the group-average representation, apply \textit{Theorem 1} across the thresholding parameter space $\mathbb{S}$, and yield the masked, binarized group-average FC; 
    \item Step 2 (Vetting Step): Compute the SNR for the masked group-average FC and investigate the SNR across all finite combinations in $\mathbb{S}$. Compute the the weak-recoverability sub-interval $I \subseteq \mathbb{S}$; 
    \item Step 3: Compute the \textit{a priori} community prominence for each individual FC; note that this prominence can also be computed for the group-average FC. 
    \item Step 4: For each individual FC, obtain the $\vec{\tau}$ maximizing the prominence computed in Step 3. Check if $\tau_{opt}$ belongs to the weak-recoverability sub-interval $I$. Note that similar to step 3, this step can also be performed on the group-average FC.
    \end{enumerate}

\section{Application: A pipeline for thresholding functional connectomes}
The below pipeline describes the process to compute the optimal threshold for a given fMRI condition, a Schaefer granularity, and a cohort in two particular cases: 
\begin{itemize}
	\item individually driven threshold $\tau^{\gamma}_{opt}$;
	\item constant (cohort-driven) threshold $\tau^{GA}_{opt}$ where $GA$ stands for group-average
\end{itemize}
Here, we see that the parameter space $\mathbb{S}$ reduces to the line search of threshold value $\tau=[0,1]\in \mathbb{S}$.
\begin{figure*}[!ht]
	\centering
	\includegraphics[width=.78\linewidth]{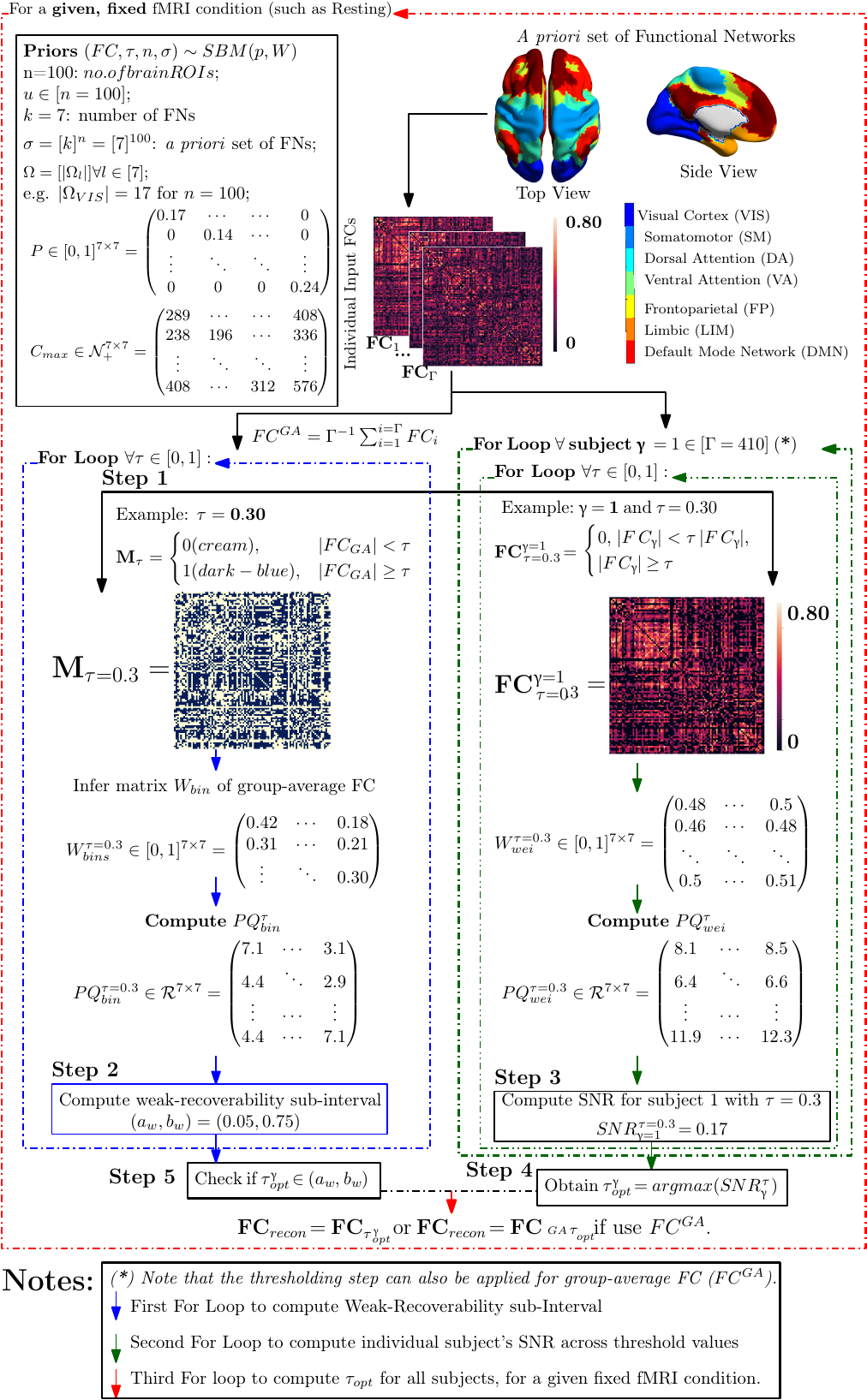}
	\caption{\small Example of the FC reconstruction routine based on the Schaefer granularity level of 100 nodes and resting-state fMRI with scanning pattern $LR$. Note that the \textit{for-loop} indicated by $(*)$ is used to find the individualized optimal threshold for each subject, $\tau^{\gamma}_{opt}$. One can substitute this \textit{for-loop} by finding the unique cohort optimal threshold, $\tau^{GA}_{opt}$, using the group-average FC.}
\end{figure*}
The pipeline contains four distinct steps:
\begin{enumerate}
	\item Step 1: For each Schaefer granularity level and task, compute the binarized (masked) group-average FC (denoted as $M^{GA}_\tau$)  using the entry-average of individual FCs (the number of individual FC is denoted as $\Gamma$)
	\item Step 2 (\textbf{Vetting Step}): 
	\begin{enumerate}
		\item 	For each threshold value $\vec{\tau}=\tau\in [0,1]$, \textbf{infer} the Stochastic Block Model (SBM) parameters to compute the Signal-to-noise ratio (SNR) of $M^{GA}_\tau$: 
		$$
		SNR[M^{GA}_\tau]=\frac{\lambda_2^2}{\lambda_1}\cbrac{[PQ]^{GA}_{bin}}
		$$
		\item Repeat this computation for all threshold values, apply \textbf{Theorem 1} to determine the weak-recoverability sub-interval $(a_w,b_w)\subsetneq \tau=[0,1]$ for the group-average FC, \textit{i.e.} $M^{GA}_\tau$ 
	\end{enumerate}
	\item Step 3: For a given individual FC and threshold value $\tau$, compute the associated thresholded FC, \textit{i.e.}, $FC^{\gamma}_\tau$, and then compute the Stochastic Block Model (SBM) parameters for $FC^{\gamma}_\tau$. Extend the usage of $SNR$ as a \textbf{mesoscopic prominence measure}: $$SNR[FC^{\gamma}_\tau]=\frac{\lambda_2^2}{\lambda_1}\cbrac{[PQ]^{\gamma}_{wei}}$$
	Analogously, we can also compute the $SNR$ for the group-average FC ($FC^{GA}$) as follows: 
	$$SNR[FC^{GA}_\tau]=\frac{\lambda_2^2}{\lambda_1}\cbrac{[PQ]^{GA}_{wei}}$$
	Repeat step 3 for all threshold values $\tau\in [0,1]$ and all individual FCs for a given fixed Schaefer parcellation and fMRI task pair;
	\item Step 4: 
	\begin{enumerate}
		\item Obtain the threshold value that maximizes SNR of the thresholded FC and the corresponding optimally reconstructed whole-brain FC; 
		$$
		\tau^{\gamma}_{opt}=\underset{\tau}{\text{argmax}}(SNR[FC^\gamma_\tau])
		$$
		Note that if the group-average FC ($FC^{GA}$) is used in \textbf{Step 3} then:
		$$
		\tau^{GA}_{opt}=\underset{\tau}{\text{argmax}}(SNR[FC^{GA}_\tau])
		$$
		\item Check if $\tau_{opt}$ is in the weak-recoverability sub-interval computed in Step 2:
		$$
		\tau_{opt}\in (a_w,b_w)
		$$
	\end{enumerate}
	Note that one needs to check the optimal threshold against the weak-recovery sub-interval, regardless of whether it is an individualized threshold ($\tau^{\gamma}_{opt}$) or a group-average one ($\tau^{GA}_{opt}$). 
\end{enumerate} 
\begin{figure*}[!ht]
	\centering
	\includegraphics[width=\linewidth]{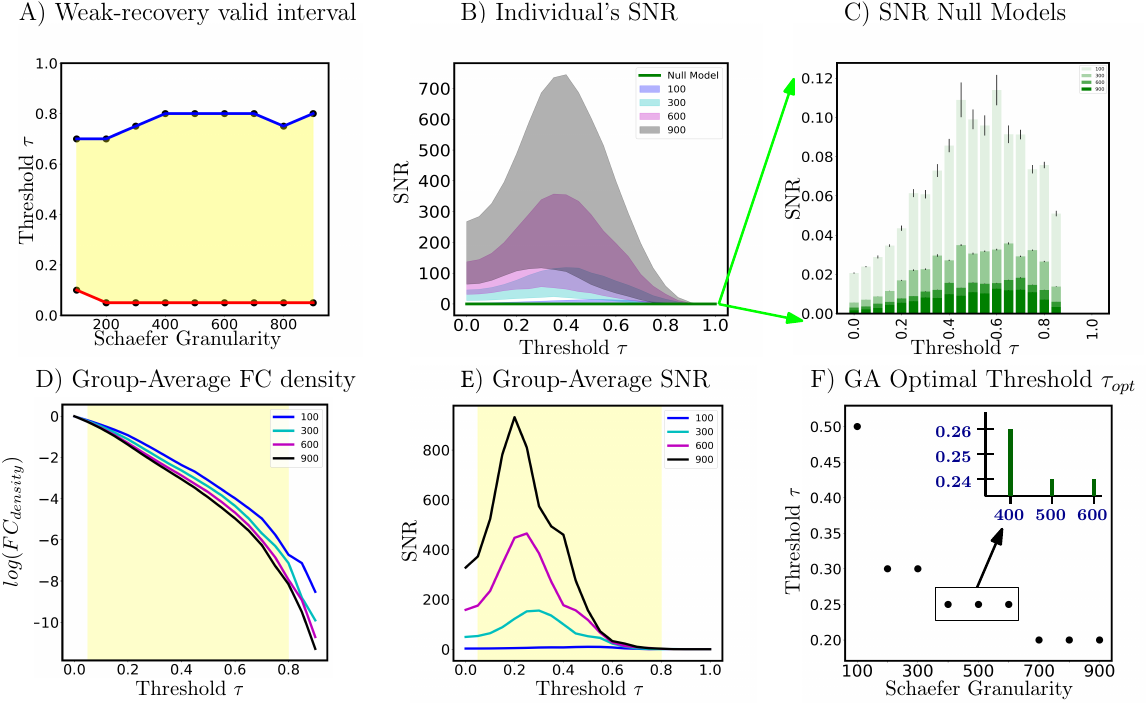}
	\caption{\small Panel (A) is the weak-recoverability sub-interval of $\tau\in (a_w,b_w) \subsetneq [0,1]$ (\textbf{Step 2}). Panel (B) is the 5- and 95- percentile of individual subjects' SNR for four distinct Schaefer parcellations $n=100,300,600,900$. Panel (C) illustrates the SNR null models. Panel (D) is the FC density, on a logarithmic scale, across the same 4 granularity levels. Panel (E) shows SNR profiles computed on group-average FCs (again, over the same granularity levels). Finally, Panel (F) reports the optimal threshold $\tau_{opt}$ computed based on the maximum SNR of group-average FCs. Note that in panels (D) and (E) the weak-recoverability sub-intervals use the maximum and minimum values for the upper and lower bound, respectively, across Schaefer parcellations.}
	\label{fig:WR_result2}
\end{figure*}	
\section{Results}

In this section, using weak recovery criteria, we investigate the level of information-theoretic prominence of an \textit{a priori} set of FNs with respect to different FCs (both group-average and individual subject levels) across a range of threshold values. Additionally, we offer deeper insights into the use of SNR as a measure of the information-theoretic prominence of this predetermined set of FNs. 

The dataset used in this paper contains 410 unrelated subjects from {the Human Connectome Project repository, Released Q3 \cite{van2012human,van2013wu}}. This includes (test and retest) sessions for resting state and seven fMRI tasks: gambling (GAM), relational (REL), social (SOC), working memory (WM), language processing (LANG), emotion (EMOT), and motor (MOT). Whole-brain FCs estimated from this fMRI dataset include 9 distinct Schaefer granularity levels that parcellate the cortical regions into $n=100$ to $n=900$ nodes, with a $100$ nodes increment for each parcellation. The functional communities evaluated in this framework include seven cortical resting state FNs from \cite{yeo2011organization}: visual (VIS), somatomotor (SM), dorsal attention (DA), ventral attention (VA), frontoparietal (FP), limbic (LIM), default mode (DMN). Each Schaefer granularity has a corresponding Yeo's FN parcellation. Additional details about the dataset are available in Supplementary Information.
\subsection{Weak-recoverability sub-interval $(a_w,b_w)$}
Based on panel (A) of figure \ref{fig:WR_result2}, we see that for most Schaefer granularity levels (except for $n=100$), the lower and upper bound of theoretically guaranteed sub-interval of weak-recovery stay fairly stable: $\tau \in[0.05,0.8]$. The lower bound $a_w$ stabilizes faster than the upper bound $b_w$, across Schaefer parcellations.  Except for the low-resolution parcellation $n=100$, the weak-recovery valid range is relatively stable and parcellation-independent. This implies that the information-theoretic relevance of an \textit{a priori} set of FNs is, to some extent, parcellation-free. In other words, for all investigated granularity levels, the thresholded graphs are in the weak-recoverability regime, except for the complete ( $\tau \in [0,0.05)$) or empty ($\tau \in (0.8,1]$) graph extremes. Panel \textbf{D} of figure \ref{fig:WR_result2} shows further details on the FC density. This is rather interesting because, at those two extremes, networks will contain either too much noise (complete graphs) or too little signal (empty graph) for any highly putative partitions to be information-theoretically relevant.

\subsection{Resting State: Group-Average versus Individuals}

Based on panels (B) and (E) of figure \ref{fig:WR_result2}, it is evident that all SNR profiles (including the group average and individual levels) behave non-monotonically across the threshold range. There exists a threshold value such that SNR is maximized in the investigated range $\tau\in [0,1]$. In addition, all optimal threshold values, for both group-average and individual FCs, are within the weak-recoverability sub-interval $(a_w,b_w)$ for all investigated Schaefer granularity levels.

Secondly, we see that both group-average and individual SNR profiles scale with $n$. This is because the scaling factor $s_t$ for the Schaefer FC sequence is not constant. In other words, as the graph size gets larger, one can expect the community profile matrix $PQ$ with entries $[PQ]_{ij},\forall i,j\in[k]$ to represent the number of expected "friends" between FN $i$ and $j$ (\textit{e.g.} between DMN and LIM) to get larger numerically. Further evidence on empirical exploration of the Schaefer graph sequence degree regime is located in Supplementary Information.

Thirdly, for a fixed Schaefer granularity level, the group-average SNR peaks \textbf{higher and earlier} across the investigated threshold range than that of an individual subject. Interestingly, the topological property of \textit{connected components} for both individual and group-average FCs, across all Schaefer parcellations, also exhibit a similar trend. Specifically, according to Figure S1 (Supplementary information), individual FCs get fragmented earlier, \textit{i.e.}, the number of connected components surpasses 1 faster, compared to the corresponding group-average FCs for a fixed granularity level. Topologically and numerically speaking, averaging FC entries across the subject domain damps down the individual fingerprints presented as high-magnitude Pearson correlation values in FCs. This results in magnitude-wise smaller functional connectivity entries, which get annihilated by smaller threshold value $\tau$. On the other hand, using the same analogy, one can see that it takes a higher threshold value for individual FC entries to be annihilated. 

\begin{figure*}[!ht]
	\centering
	\includegraphics[width=\linewidth]{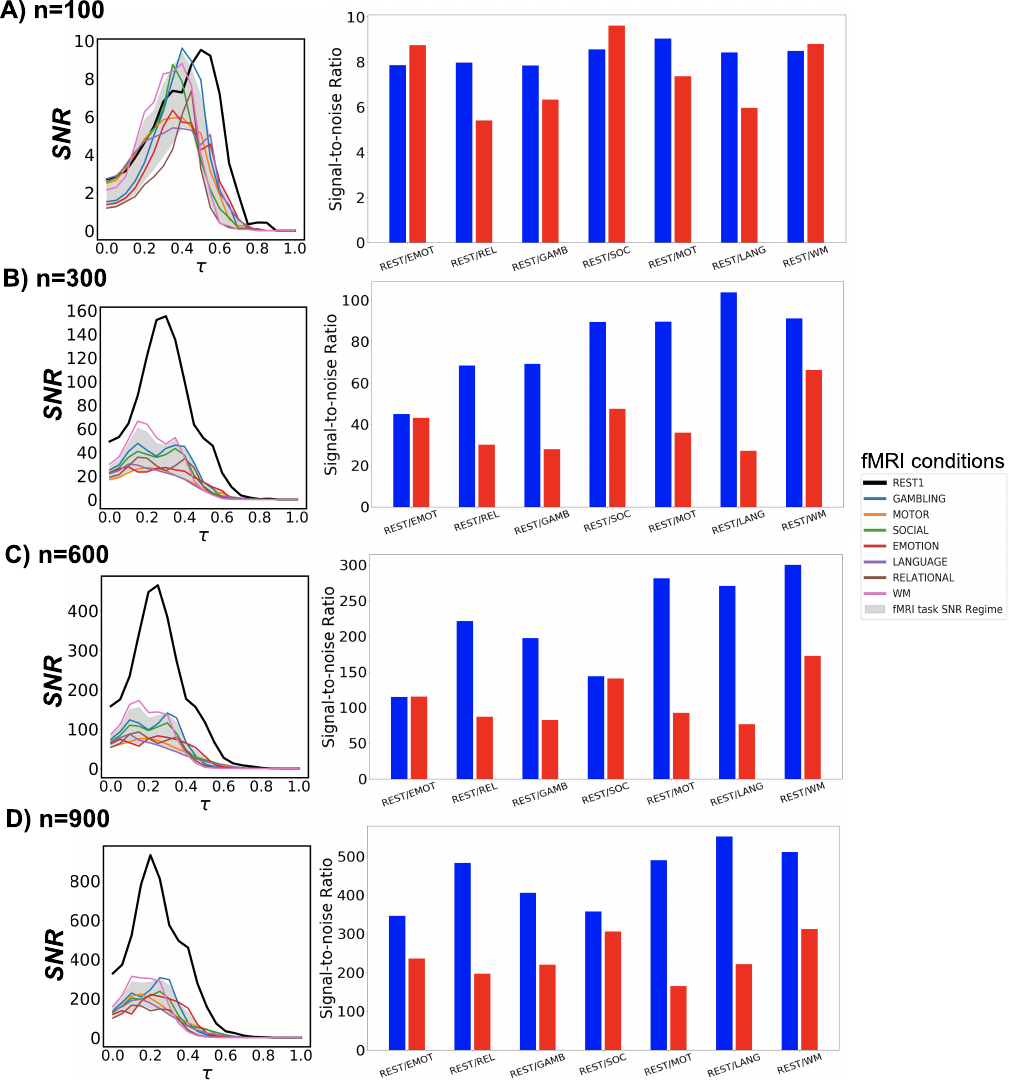}
	\caption{\small \textbf{fMRI task and rest SNR profiles.} For each row of panels, the left one represents SNR behavior across threshold range $\tau\in[0,1]$ using the \textbf{maximum} scanning length for all fMRI tasks and rest. The gray shade represents the $5$- to $95$- percentile of SNR task regime across all fMRI tasks and the entire threshold range. In the right panel, SNR profiles for task- and resting-state fMRI are computed using 166 time points corresponding to the scanning length of the shortest scanned task, \textit{i.e.}, EMOTION. Results are for 4 Schaefer parcellation levels: $n=[100,300,600,900]$.}
	\label{fig:WR_task}
\end{figure*}

\subsection{Individualized optimal thresholds}
\begin{figure}[ht!]
    \centering
    \includegraphics[width=.7\linewidth]{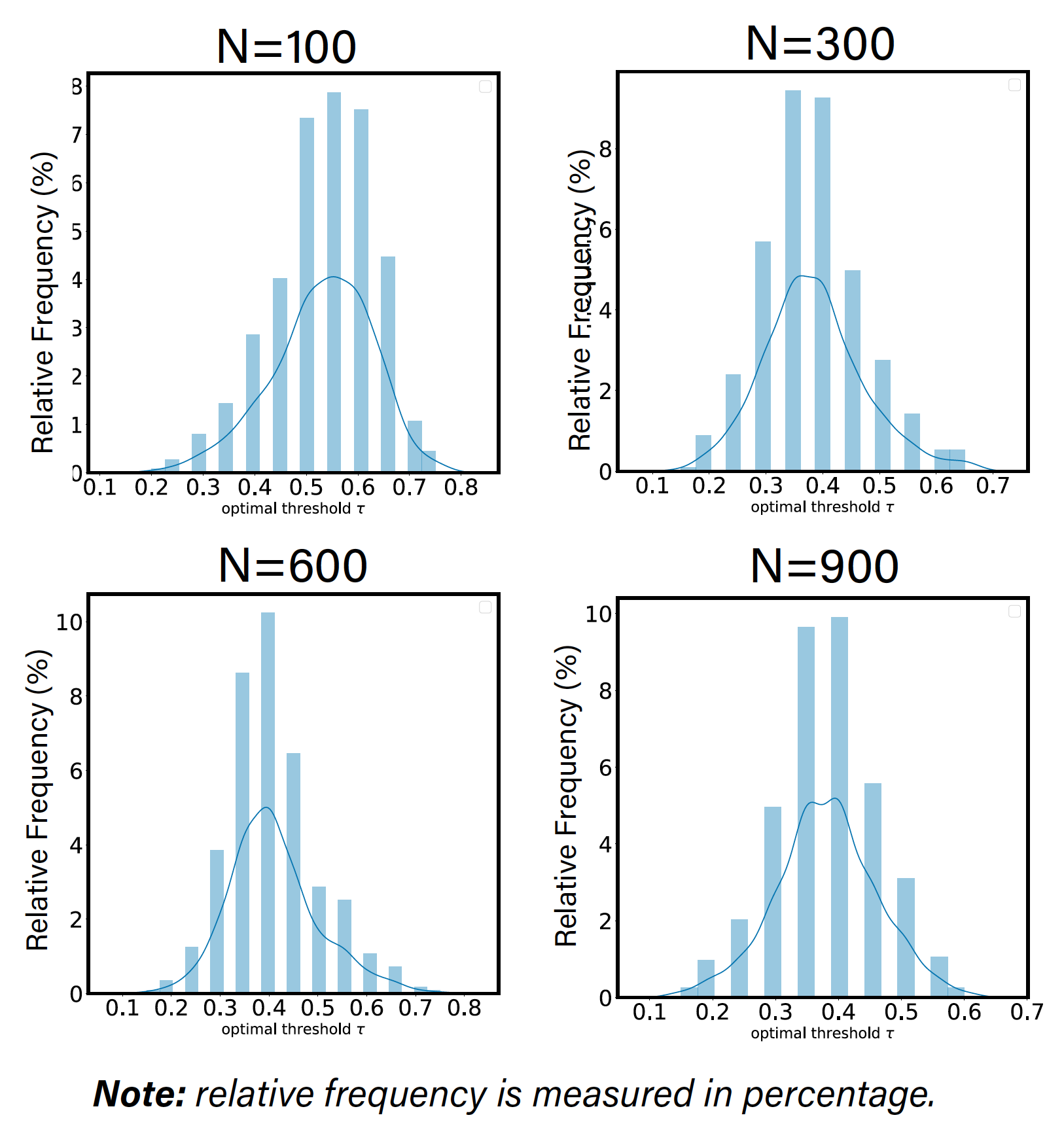}
    \caption{Individualized optimal threshold is derived using the SNR behavior of each individual for 4 distinct Schaefer's parcellations $n=\{ 100, 300, 600, 900\}$.}
    \label{fig:indiv_opt_tau}
\end{figure}

As one can observe from figure \ref{fig:indiv_opt_tau}, the individualized optimal threshold varies across different individuals, which demonstrates strong evidence of the existence of FN functional fingerprint across subjects. In addition, the average of these individualized thresholds, for a given parcellation granularity, is roughly equal to the group-average optimal threshold.  

\subsection{Group-average: Resting State vs. fMRI Task Analysis}

Next, we investigate the prominence of Yeo's resting state networks with respect to different fMRI conditions, including 7 tasks and the resting state, through SNR measures using group-average FCs across all Schaefer granularity levels and the entire threshold range. Using the resting state SNR profile as a baseline, we compare all task responses in these two scenarios: 
\begin{itemize}
	\item constructing FCs with the maximum number of time points available for each fMRI condition;
	\item for all fMRI conditions, constructing FCs using 166 time points, which correspond to the number of time points associated with the EMOTION task that is also the minimum across all conditions.
\end{itemize} 

Firstly, in both scenarios,  the \textbf{maximum} SNR values for all examined tasks surpass the hard threshold $SNR=1$ for weak recoverability. Moreover, the optimal threshold $\tau_{opt}$ consistently falls within the range $(a_w,b_w)$. Trivially, resting state SNR dominates all available tasks across all parcellation levels. This is expected because the selected set of FNs are Yeo's resting state networks. Secondly, working memory (WM) fMRI responds fairly consistently across all granularity levels in both scenarios. From an information-theoretic perspective, EMOTION is the most similar task to the resting state, with respect to Yeo's resting state networks. 
	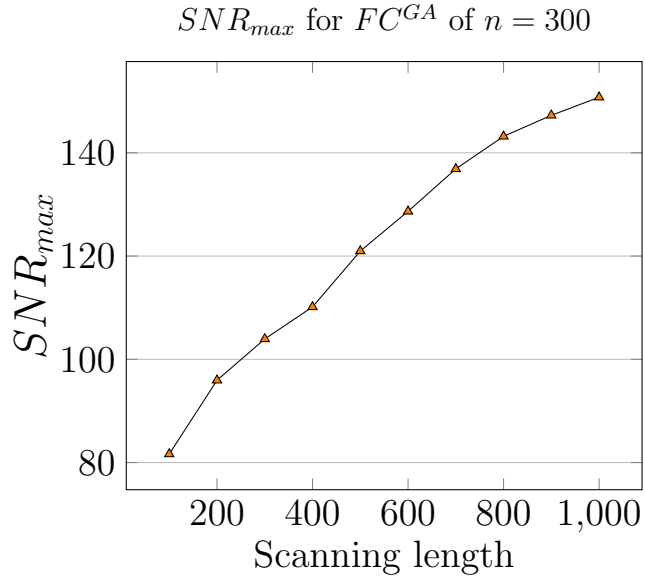
\begin{figure}[!ht]
	\centering
	\begin{adjustbox}{max width=\columnwidth}
		\begin{tikzpicture}
		\begin{groupplot}[
		%		legend columns=6,
		%		legend entries={10MB, 100MB, 1GB, 10GB, 100GB, 1TB},
		%		footnotesize,
		%		legend to name=LegendCompTime,
		every x tick label/.append style={font=\large},
		every y tick label/.append style={font=\large},
		y label style={font=\Large},
		x label style={font=\large},
		group style={
			group size=1 by 1,
			xlabels at=edge bottom,
			ylabels at=edge left}]
		\nextgroupplot[title={$SNR_{max}$ for $FC^{GA}$ of $n=300$}, ylabel=$SNR_{max}$, xlabel=Scanning length, log ticks with fixed point,ymajorgrids]
		\addplot[mark=triangle*, orange, draw=black] coordinates {
			(100, 81.65110826)
			(200, 95.94080119)
			(300, 103.94047181) 
			(400, 110.14972893) 
			(500, 120.97147901)
			(600, 128.66419775)
			(700, 136.88361667)
			(800, 143.19315099) 
			(900, 147.26241098) 
			(1000, 150.7638001) 
		};
		\end{groupplot}
		\end{tikzpicture}
	\end{adjustbox}
	%	\begin{adjustbox}{max width=\columnwidth,scale={1.0}}
	%		\ref{fig:increasing_length_SNR}
	%	\end{adjustbox}
	\caption{Maximum $SNR$ computed for Resting state of Schaefer Group-average FC with $n=300$ for increasing scanning lengths, starting at 100 to 1000 time points, increments of 100 each.}
	\label{fig:increasing_length_SNR}
\end{figure}

Thirdly, in the maximum-timepoint case, with the exception of $n=100$ parcellation, the SNR profiles for most tasks are roughly half the magnitude of the resting-state SNR. Furthermore, for all examined Schaefer parcellations, group-average task FCs appear to reach their SNR peak earlier than the resting-state counterpart. Further details are indicated in figure \ref{fig:WR_task} - Panel \textbf{A}.

In the second scenario when the minimum number of time points is used across all fMRI conditions, the gap in SNR magnitude between the resting state and each task condition is significantly narrowed, yet the SNR during rest still exceeds those during tasks. Further details are indicated in figure \ref{fig:WR_task} - Panel \textbf{B}. 
% Note that the gray shaded area indicates the 5- and 95- percentile of SNR responses among all fMRI tasks.

\subsection{The SNR-driven inequality}
It is important to check if SNRs are robust against randomness, \textit{i.e.}, whether they are a valid factor in deciding the threshold. To do so, we randomly shuffle Yeo's resting state networks and recompute the SNR response. We repeat the random shuffling procedure 100 times and record the results for all nine group-average FC induced by the nine Schaefer parcellations, each of which is under the $REST_1$ condition with scanning pattern $LR$. Results for $RL$ pattern are available in Supplementary Information.

For every fixed Schaefer parcellation granularity, the null model SNR profiles are uniformly lower than those of all subjects across the entire thresholding range. Furthermore, the null model values do not exceed the hard threshold imposed by weak recovery criteria, \textit{i.e.} $SNR=1$. This observation holds true across all investigated Schaefer parcellations, as seen in panel \textbf{F} of figure \ref{fig:WR_result2}. Interestingly, the SNR gets uniformly smaller as Schaefer parcellation granularity increases, as seen also in Panel \textbf{F}.

Collectively, given the SNR results obtained at rest, under task conditions, and null models, we empirically form an inequality relation between resting state and task fMRI-induced FCs in terms of SNR response and the corresponding level of prominence of Yeo's resting state networks across different fMRI conditions: 
\begin{equation}
0<SNR_{null}<SNR_{task}<SNR_{rest}
\end{equation}
This general order of \textit{SNR} response is observed at the threshold $\tau$ that maximizes the objective function $SNR$ by the weak-recovery criteria.  At such optimal threshold values, all SNR profiles for task fMRI are in the weak recoverability region while still smaller in magnitude than that at the resting state. Together, these inequalities constitute an empirical lower-bound and upper-bound for $SNR_{task}$, at least for all the tasks investigated in our study. 
\subsection{Maximum SNR and threshold relationship}
As the granularity of Schaefer parcellation gets finer, the corresponding group-average SNR profiles get larger due to the natural scaling of the community profile matrix $PQ$. This observation applies to the majority of the threshold range. Moreover, per figure \ref{fig:WR_result2} Panel \textbf{F}, we see that optimal thresholds, \textit{e.g.} $\tau_{opt}$, tend to decrease as the granularity level increases, which suggests that larger Schaefer FCs do not need to be thresholded as much. Another interesting observation is that with the exceptions of $n=\cbrac{100,200,900}$, all other investigated granularity levels accept a very stable optimal threshold $\tau^{GA}_{opt}=0.25$. Being a computation pipeline that relies on discretized line search on threshold $\tau$ (of increments 0.05 for $\tau=[0,1]$), yielding this level of consistence of optimal value is unexpected. 
\begin{figure*}[!ht]
	\centering
	\includegraphics[width=\linewidth]{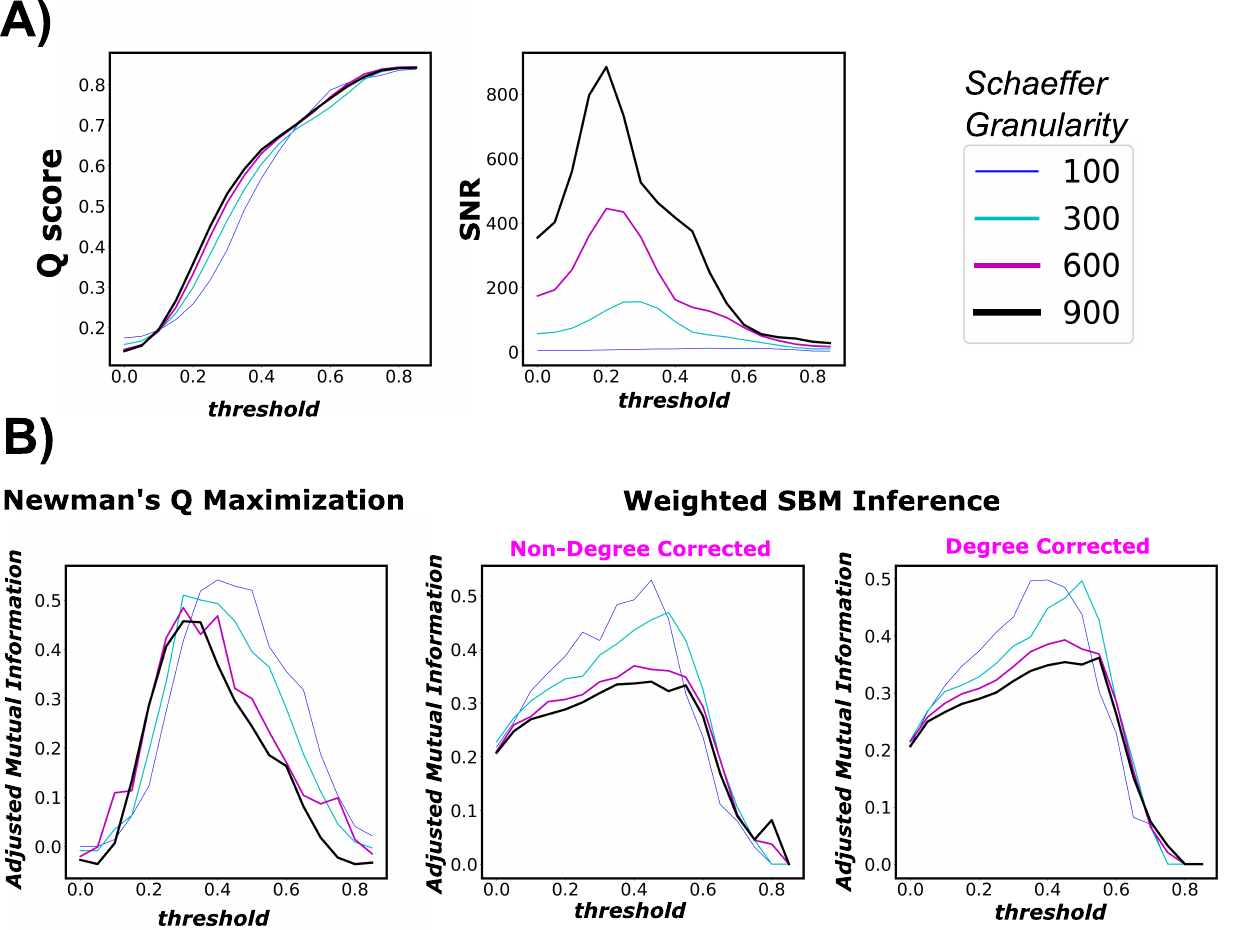}
	\caption{\small Panel (A) - left figure represents the modularity score of a thresholded group-average FC across threshold range $\tau\in[0,0.85]$. Panel (A) - right figure reports the normalized mutual information between the inferred partition (using $Q$-score maximization heuristics) and Yeo's FN partition. The same order goes to Panel (B).  Panel (B) represents the results of the SNR approach. Note that the full threshold range is not necessary because in the sub-interval $\tau\in[0.9,1.0]$, the thresholded graph is almost (if not) empty. The displayed result is for the group-average FCs, over four Schaefer granularity levels $n=[100,300,600,900]$.}
	\label{fig:WR_backtest}
\end{figure*}	
\subsection{Validation: Highly-putative partition back-test}
The theory of weak recovery and its extended usage proposed here allowed us to argue for the relevance of using SNR as a measure that, {via thresholding as a specific use case, guides the estimation of well-defined functional connectivity given the mapping of an \textit{a priori} set of FNs.} {
In this section, our goals are two-fold:
\begin{enumerate}
    \item validate our framework by solving the "forward" problem (e.g., community detection) and compare the detected partition with the ground truth;
    \item benchmark different community detection algorithms to solidify the above validation.
\end{enumerate}
}

Specifically, we juxtapose SNR as a guiding measure against objective-function community detection methods. One such method is Newman's $Q$-score maximization \cite{newman2004finding,newman2006finding,newman2006modularity}. {Here, instead of using $Q$-score as an objective function for community detection, we use it as a guiding measure to investigate its behavior across the tested threshold range $[0,1]$.} In broad strokes, the $Q$ score, \textit{modularity score}, measures the statistical difference between a network and its corresponding null model with similar topological properties such as the degree sequence. It can be computed as follows:
$$
Q=\sum_{u,v} (A_{uv} -\alpha P_{uv})\delta(\sigma_u,\sigma_v)
$$
where $\delta(\bullet,\bullet)$ and $\alpha$ are the Kronecker delta and tuning parameter (defaulted at $\a=1$), respectively. In network neuroscience, the majority of studies examining mesoscopic structures of brain functions heavily leverage the maximization of $Q$ score, which unravels predominantly assortative communities, \textit{i.e.}, mesoscopic structures with denser internal edge density than the external one \cite{sporns2013network,sporns2016modular}. SBM inference methods like Weighted SBM Inference, in principle, uncover a more diverse type of communities beyond assortative ones, such as dis-assortative and core-periphery communities \cite{betzel2018diversity}. Because of such distinct differences in principle between the two types of approaches, $Q$ score would provide a good benchmark for comparing the robustness of SNR against various community detection approaches. Note that for Weighted SBM inference, we assume a Poisson distribution for the weighted graph \cite{karrer2011stochastic}. Although other model assumptions are possible, our goal in this paper is not to select the most fitting model assumption but rather to investigate the differences in the communities detected using two 
% physiologically 
theoretically different approaches. 
In other words, we are not looking to see if $Q$ score or $SNR$ picks up the exact threshold where the inferred partition is information-theoretically aligned with Yeo's FNs; rather, we are interested in seeing whether each of those two measures captures the threshold interval where the two partitions agree to a relatively high degree. 
To measure the information-theoretic agreement between the inferred and ground-truth partitions, we use adjusted mutual information (AMI), which is a measure adjusted to chance. Further details on the inference method and  AMI are described in the Supplementary Information.

{Here, our evaluation criteria are as follows: there would only be a threshold sub-interval that are better aligned with the highly-putative (ground-truth) community assignment. In the forward direction, we apply 3 different community detection algorithms across the investigated threshold range $[0,1)$ and measure the corresponding AMI between the inferred and ground-truth partitions. A robust guiding measure, since "blinded" from the ground-truth partition, should have a similar behavior, compared to the AMI curve. }

Firstly, per figure \ref{fig:WR_backtest} right panels, we see that both community detection methods, namely Newman's $Q$-score Maximization and Weighted SBM Inference, yield very similar trends. Specifically, both AMI profiles go up and down crossing the threshold range. Further, AMI gets smaller as $n$ gets larger, which is expected for graphs with increasing numbers of nodes. Interestingly, the threshold values that maximize the AMI for Newman's $Q$-score Maximization tend to shift left as $n$ increases. We see this particular behavior with $SNR$ in the earlier result section (Panel F of figure \ref{fig:WR_result2}). Secondly, the $Q$ score keeps a fairly steady rise in magnitude across the threshold range. Further, it does not appear that the $Q$ score is parcellation dependent; this is expected because the measure is normalized by $2m$. Moreover, $Q$ score peaks and plateaus at a very high threshold range $\tau\in [0.6,0.8]$. In that range, the thresholded FC is highly fragmented (Figure S1) with extremely low edge density \ref{fig:WR_result2} and ceases to retain interesting topological insights for further analysis. Lastly, we see that $SNR$ driven curves, with an \textit{a priori} set of FNs, behave very similar to AMI profiles of both objective-function approaches. On the other hand, the $Q$ score keeps rising across the threshold range and starts plateauing towards the end, which indicates failure at picking a threshold useful for an \textit{a priori} partition such as Yeo's FNs. Collectively, our results show, once again, that $SNR$ computation on weighted, thresholded FCs provides excellent guidance for reconstructing a graph with the most information-theoretic relevance to a particular fixed set of FNs.  
	\section{Discussion}
%Intro
In recent years, the network neuroscience field has been striving forward with many exciting discoveries that are becoming more and more relevant to clinical applications and personal medicine. In network neuroscience, this urges the need to improve a popular proxy of brain function, namely functional connectivity. Having the most proper, state-of-the-art mathematical representation of functional brain circuits allows for more accurate and confident positioning of research endeavors. In this work, we put forth a simple framework that allows improving the mathematical representation of brain functions given the use of an \textit{a priori} set of functional networks. This framework also doubles as a clear evaluation tool for any specific combination of FC parcellation, FN partition, and edge pruning techniques applied to a large-scale brain dataset, thereby streamlining the complex yet crucial studies in network neuroscience.

Thresholding, which is an edge pruning technique used in post-FC processing, is seldom challenged as a standard practice that eliminates, albeit arbitrarily, statistically spurious edges. Since an increasing body of clinical research now involves FC thresholding in the data construction pipeline, careful scrutiny of thresholding is therefore imperative. We conclude that there is no single constant threshold value that is optimal across different parcellation granularity levels, such as the Schaefer ones. In particular, from coarser to finer Schaefer granularities, the optimal threshold value decreases. This result is partially observable in the behavior of matrix $W$ across all studied Schaefer parcellations. 
According to figure \ref{fig:WR_backtest}, we see that for a fixed threshold value, as Schaefer granularity increases, Yeo's functional networks behave more in an assortative manner, \textit{i.e.}, denser internal edge density and sparser external one. We see that through a brighter diagonal and a darker off-diagonal regime of matrix $W$, across Schaefer parcellations with fixed threshold value $\tau$. 
Information-theoretically, it means that a larger graph (in size) tends to contain more relevant information to unravel the ground-truth partition (in our study, seven Yeo's resting state networks); hence, we do not need to threshold the FCs as deeply as the lower granularity parcellations such as $n=100$. This result also suggests that FC size is proportional to the level of prominence, or fitness, of the \textit{a priori} set of FNs. Nonetheless, the exact limit of this behavior when granularity tends toward infinity is unknown, \textit{e.g.}, whether the optimal thresholding value will reach a plateau even if the granularity increases. {Despite this uncertainty, we showed that for the majority of granularity levels between 100 and 900, the commonly used threshold of $\tau = 0.25$ is a valid choice for group-average topology.}

%\textbf{Addressing the dawn of individualized parcellations}
Moreover, when using SNR as a goodness-of-fit measure while fitting an \textit{a priori} set of FNs onto the FC, while no significant differences are observed between resting state and task conditions for the low-resolution parcellation (n=100), distinct differences emerge at higher resolutions. There are two ways of interpreting this result: i) \textit{a priori} FNs exhibit a poorer fit during rest compared to task states; ii) there is an intrinsic shift of functional network dynamics at the individual level between the resting state and the task condition. Furthermore, there is also strong evidence suggesting a wide variance in the individualized thresholds across all Schaefer parcellation granularity levels. In the same vein, our results also support the concept of \textit{individualized parcellation} suggested by the work of Salehi and colleagues \cite{salehi2018there}. While intuitive and insightful, individualized parcellation across subjects and tasks remains computationally expensive. To that end, our work offers a well-defined tool to examine the level of relevance a particular set of functional networks exhibits when mapped onto individual FCs under different conditions. In simpler terms, it allows us to, for the first time, quantify the individual differences (through information-theoretic gap) when the same atlas is mapped across cohort and/or task domains. This paves the way for alternative frameworks that build upon our work, potentially leading to task-dependent or subject-dependent parcellation methods beyond that proposed in  \cite{salehi2018there}. 

%Limitation and future direction
Our work also extends the usage of the weak-recovery theorem by leveraging SNR as a goodness-of-fit measure. Specifically, our results suggest that for the majority of threshold values, the masked binarized FCs are in the regime of week recovery. However, an open question remains: when parcellated by the Schaefer atlas for a fixed individual and an fMRI condition, is the sequence of FCs in the exact recovery regime? Future studies are needed to address this information-theoretic gap between weak and exact recoverability requirements that is reflected by two measures: SNR for weak recovery and Chernoff-Hellinger distance for exact recovery. Although exact recovery is a stronger requirement, if the Schaefer graph sequence falls within the exact recovery degree regime, the mutual information between the inferred partition (through network inference and objective-based community detection methods) and the ground-truth one (\textit{e.g.}, Yeo's parcellations) will be theoretically higher. 
Furthermore, future work should address limitations in the fMRI voxel resolution, both spatial and temporal, and those in the corresponding Schaefer parcellations, particularly their impact on the SNR when fitting Yeo’s functional networks. Specifically, further investigation should be done on the effect of voxel sizes (\textit{e.g.}, 2 mm isotropic for the HCP data set \cite{van2012human}) and the repetition time (\textit{e.g.}, 720 ms for the HCP data set \cite{van2012human}).

%\textbf{Addressing our approach to contribute in this avenue.}
Lastly, our findings highlight two important points for brain connectomics research. First, because of the existence of individual brain fingerprints \cite{finn2015functional,amico2018quest}, we need to pay extra attention when applying a common, fixed atlas to individual FCs. Secondly, we show that thresholding FC matrices is not only an intuitive step during FC post-processing (\textit{e.g.}, to eliminate statistically spurious edges) but also a necessary one if we would like to use such FCs, coupled with an \textit{a priori} set of FNs, to support any research endeavor in brain connectomics. These results suggest a promising new direction: individualized and task-dependent parcellation methods as an alternative to fixed atlases like Yeo's.  {
{This opens up new directions for precision medicine, particularly targeting individual-specific neurological and neuropsychiatric disorders.}
}

\newpage

\section*{Author Contributions}
D.D.-T.: Conceptualization; Formal analysis; Investigation; Methodology; Writing original draft. J.C-P., A.D.K: Investigation; Methodology. N.N: Investigation; Formal Analysis; Visualization; Writing original draft. S.M., J.C., J.B., F.X., S.G., Y.Z.: Writing—review and editing. J.G., L.S.: Conceptualization; Formal analysis; Writing original draft; Project Supervision; Funding acquisition. All authors have read and agreed to the published version of the manuscript.
\section*{Funding}
JG acknowledges financial support from NIH R01EB022574 and NIH R01MH108467 and the Indiana Clinical and Translational Sciences Institute (Grant Number UL1TR001108) from the National Institutes of Health, National Center for Advancing Translational Sciences, Clinical and Translational Sciences Award. LS acknowledges financial support from the National Institutes of Health grants RF1 AG068191, R01 AG071470, U01 AG068057, and T32 AG076411, the National Science Foundation grant IIS 1837964. NN acknowledges financial support from the Erasmus Mundus Joint Master's Programme in Brain and Data Science, European Commission. DDT acknowledges financial support from Office of Naval Research N0001423WX00749 and Lawrence Livermore National Laboratory under Contract DE-AC52-07NA27344.
\section*{Data Availability Statement}
Data were provided [in part] by the Human Connectome Project, WU-Minn Consortium (Principal Investigators: David Van Essen and Kamil Ugurbil; 1U54MH091657) funded by the 16 NIH Institutes and Centers that support the NIH Blueprint for Neuroscience Research; and by the McDonnell Center for Systems Neuroscience at Washington University. This data used in this study are freely available on the HCP website (https://www.humanconnectome.org, accessed on 1 September 2021). The release Q3 from the HCP data with resting state and seven fMRI tasks and Glasser parcellation was used, and users must apply for permission to access the data. 
\section*{Acknowledgments}
Data were provided (in part) by the Human Connectome Project, WU-Minn Consortium (principal investigators: David Van Essen and Kamil Ugurbil; 1U54MH091657) funded by the 16 NIH Institutes and Centers that support the NIH Blueprint for Neuroscience Research; and by the McDonnell Center for Systems Neuroscience at Washington University. We aslo thank Dr. Emmanuel Abbe for his valuable comments

\section*{Author Declaration}
The authors declare no conflict of interest.

\newpage

\Large{\textbf{References}}\\
\bibliographystyle{abbrv}
\bibliography{SBM}

\end{document}

\title[A principled framework of functional connectome thresholding]{Supplementary information: A principled framework to assess the information-theoretic fitness of brain functional sub-circuits}

\author{
    Duy Duong-Tran$^{1,2,*,\ddagger}$, 
    Nghi Nguyen$^{3, \ddagger}$, 
    Shizhuo Mu$^{1}$, 
    Jiong Chen$^{1}$, 
    Jingxuan Bao$^{1}$, 
    Frederick Xu$^{1}$, 
    Sumita Garai$^{1}$, 
    Jose Cadena-Pico$^{4}$, 
    Alan David Kaplan$^{5}$, 
    Tianlong Chen$^{6}$,
    Yize Zhao$^{7}$,
    Li Shen$^{1,\dagger}$, 
    and Joaqu\'{i}n Go\~{n}i$^{8,9,10,\dagger}$
}

\address{
$^{1}$ \quad Department of Biostatistics, Epidemiology, and Informatics (DBEI), Perelman School of Medicine, University of Pennsylvania, Philadelphia, PA, USA\\ 
    $^{2}$ \quad Department of Mathematics, United States Naval Academy, Annapolis, MD, USA\\
    $^{3}$ \quad Gonda Multidisciplinary Brain Research Center, Bar-Ilan University, Ramat Gan, Israel\\
    $^{4}$ \quad Machine Learning Group, Lawrence Livermore National Laboratory, Livermore, CA, USA\\
    $^{5}$ \quad Computational Engineering Division, Lawrence Livermore National Laboratory, Livermore, CA, USA\\
    $^{6}$ \quad Department of Computer Science, The University of North Carolina at Chapel Hill \\
    $^{7}$ \quad School of Public Health, Yale University, New Heaven, CT, USA   \\
    $^{8}$ \quad School of Industrial Engineering, Purdue University, West Lafayette, IN, USA\\
    $^{9}$ \quad Purdue Institute for Integrative Neuroscience, Purdue University, West Lafayette, IN, USA\\ 
    $^{10}$ \quad Weldon School of Biomedical Engineering, Purdue University, West Lafayette, IN, USA \\
* Correspondence to: duongtra@usna.edu \\
$\ddagger$ Equal contribution \\
$\dagger$ Co-supervising Authors}

The purpose of this document is to elaborate on the machinery of the morphospace and other aspects such as the dataset and brain atlas used to analyze the data. The aim is to provide further analytic results in conjunction with those already presented in the main paper.
\renewcommand{\thefigure}{S\arabic{figure}}
\setcounter{figure}{0}

\section*{Schaefer Sequence $\cbrac{G_{t_\ell}}$ Topology - Resting State Analysis}
\subsection*{Number of Connected Components}
In this section, we investigate the topological features of the Schaefer FC graph sequence across the entire threshold period $\t\in [0,1]$. Specifically, we examine the number of components across the threshold range and Schaefer granularity levels.

We use all nine available Schaefer parcellations with $n=[100,200,...,900]$ and their corresponding mappings of Yeo's seven resting state networks \cite{yeo2011organization} for each granularity level. Besides the individual level FC (denoted as $FC^\gamma$), the group-average FC (denoted as $FC^{GA}$) is computed using the entry-wise mean across the individual FCs (denoted as $FC$):
\[
FC^{GA}=\frac{\sum_{\gamma=1}^{\Gamma} FC^\gamma}{\Gamma}
\]
where $\Gamma$ denotes the number of subjects and $\gamma\in[\Gamma]$. 

Specifically, for each Schaefer granularity and threshold combination, we compute the number of components for each individual and group-average FC.
\begin{figure*}[ht]
	\centering
	\includegraphics[width=\linewidth]{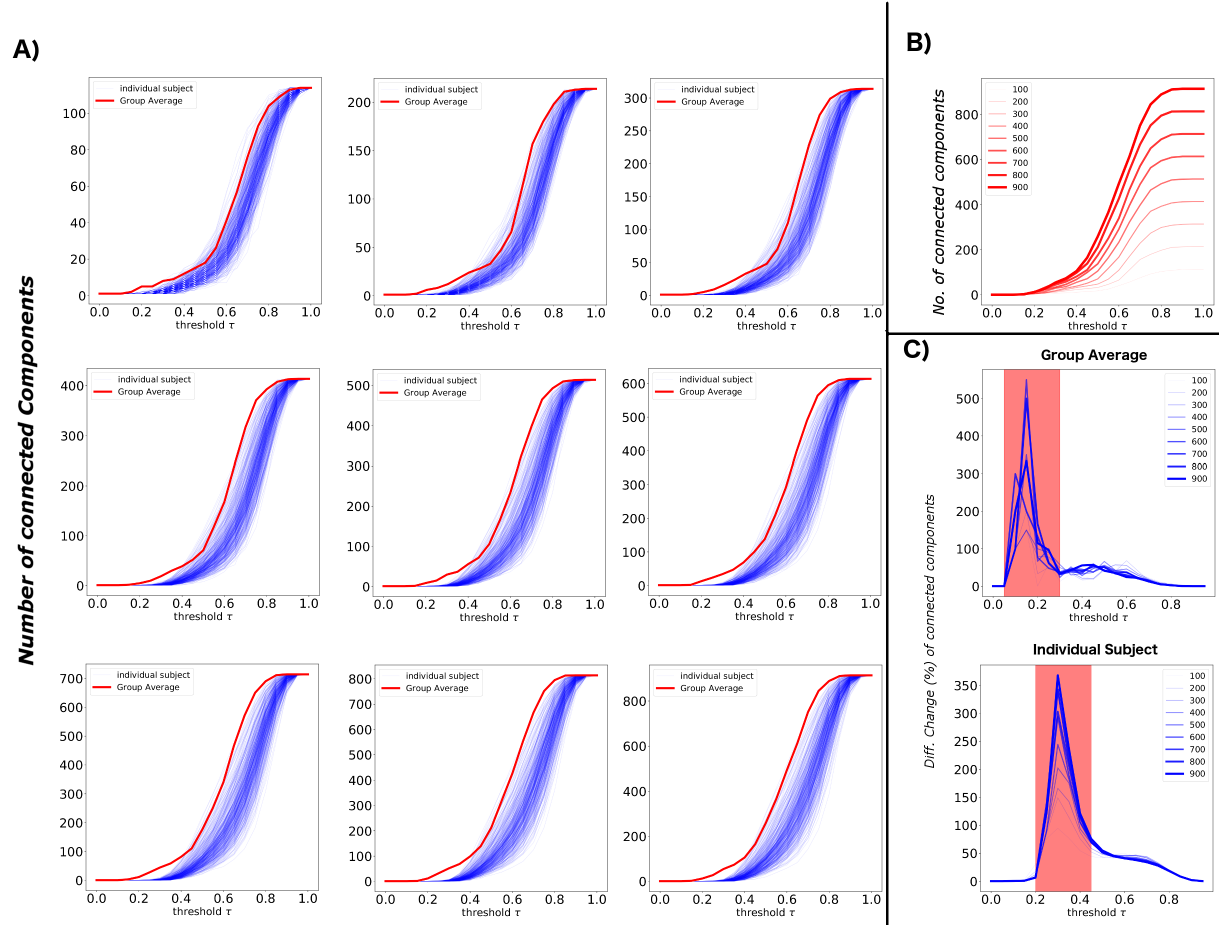}
	\caption{\small Panel (A) represents the number of connected components for each Schaefer parcellation (from 100 to 900 nodes with an increment of 100 nodes each time), across the pre-defined thresholding range $\t\in [0,1]$. Panel (B) represents the overlap in the number of components of the group-average FC for each Schaefer parcellation. Panel (C) shows the differential change (in \%) between two consecutive numbers of component statistics across $\t$ for group-average FCs (\textit{top}) and the \textbf{mean} of individual subject FCs (\textit{bottom}).}
	\label{fig:no_of_comp}
\end{figure*}

To study this characteristic, we use resting state fMRI data, \textit{e.g.}, \textit{rfMRI}. Without loss of generality, we select the first resting scan, \textit{i.e.}, \textit{REST}$_1$, with phase encoding \textit{LR}. It is important to note that the connectivity (computed as the number of connected components) of the thresholded FC (where the absolute values of functional edges are set to zero - only applying step (a) above) is analogous to its binarized thresholded counterpart (where the surviving functional edges are set to one - applying both step (a) and (b) for any given threshold and Schaefer parcellation choice). The number of components is computed using the Python package \url{networkx} after converting the FC matrix to a graph object.

Firstly, for all considered Schaefer parcellations, we observe that the group-average FC fragments (\textit{e.g.}, splits into more than one connected component) earlier than the individual subjects' FCs. This is because the normalization of functional edges across the cohort domain neutralizes individual differences and zeroes out relatively faster across the thresholding range (Panel A). Moreover, it is also expected that the group-average number of connected components increases proportionally with the parcellation sizes, and for a fixed threshold value, the number of connected components in a coarser parcellation is always smaller than in a finer one (Panel B).

We also compute the differential change $\Delta\mathbb{C}$ (in percentage) between two consecutive $\mathbb{C}$s across the threshold range as follows:
$$
\Delta \mathbb{C}_l (\%)= \frac{|\mathbb{C}_{l+1}-\mathbb{C}_{l}|}{\mathbb{C}_l} \times 100 
$$

where $l$ is indexed over the threshold range. We observe that both the group-average (Panel C-\textit{top}) and individual level (Panel C-\textit{bottom}) show an empirical phase transition in the number of connected components. This phase transition occurs in the sub-intervals $(0.05,0.30)$ and $(0.20,0.45)$ for the group-average and individual levels, respectively. Although there is a numerical overlap between the two phases, the group-average transitions earlier than the individual one.

\subsection*{FN-Differential Identifiability $\mathbb{I}_{diff}$ and Empirical Schaefer Degree Regime}
In this section, we investigate the behavior of the matrix $W_{bin}$ of group-average FCs, using Yeo's 7 resting state networks \cite{yeo2011organization}. To make some empirical observations about the Schaefer FC sequence degree regime, we examine the group-average masked FC, $M^{GA}$, across all nine granularity levels and the threshold interval $\t\in [0,1]$ with an increment of $0.05$. The reason we look only at the masked (binarized) FCs is because 

\begin{figure*}[!ht]
	\centering
	\subfloat[FN-Differential Identifiability $\mathbb{I}_{diff}^{FNs}$ score for Language fMRI task ($LR$ scanning pattern).]{\includegraphics[width=.45\columnwidth]{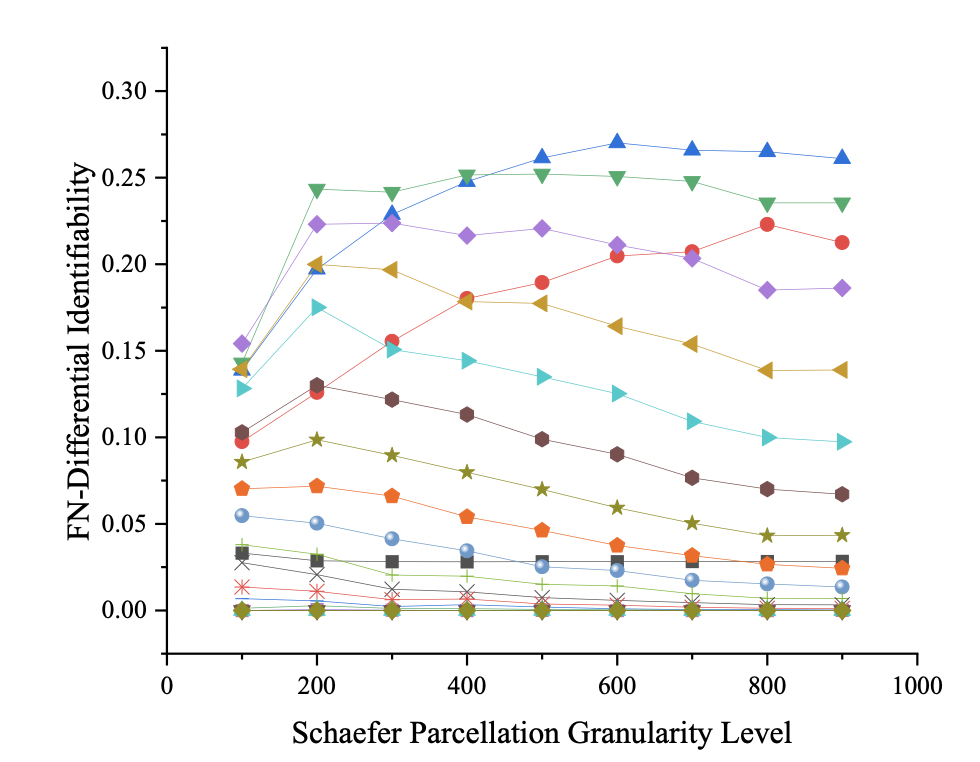}}
	\hspace{2mm}
	\subfloat[FN-Differential Identifiability $\mathbb{I}_{diff}^{FNs}$ for resting state ($LR$ scanning pattern).]{\includegraphics[width=.45\columnwidth]{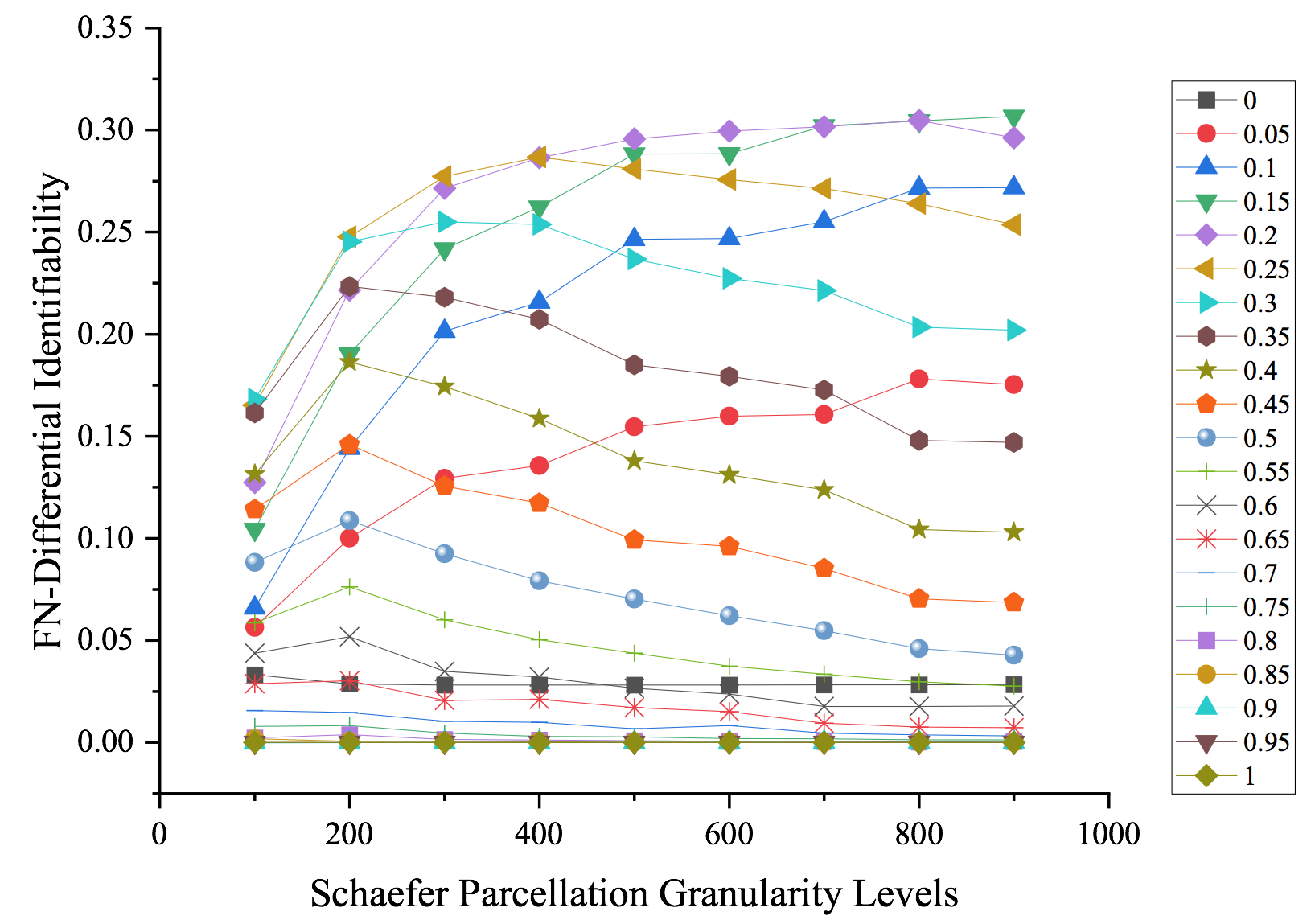}}
	\caption{FN-Differential Identifiability $\mathbb{I}_{diff}$ for resting state and one particular fMRI task across all threshold and Schaefer Granularity Level combinations.}
	\label{matrixIdiff}
\end{figure*}

\begin{itemize}
	\item The Sandon et al. \cite{abbe2017community} theorem on weak-recovery is written for binary graphs. Hence, the recoverability requirement on the degree-regime is only applied to the binary scaffold.
	\item We see that looking at weighted graphs is not appropriate in this case as the row (or column) sum of the FC matrix would yield the connectivity strength of a node, not its degree. 
\end{itemize}

Here, we investigate the empirical degree regime of the Schaefer graph sequence based on the behavior of $W_{bin}$. For all studied Schaefer granularity levels and threshold combinations, to infer $W_{bin}$, we simply use the maximum likelihood rule as mentioned in the main text. Recall that matrix $W_{bin}=[w_{ij}]$, where $w_{ij}$ contains the probability that a node $u$ in community $i$ is connected (\textit{e.g.}, $a_{uv}=1$) or not-connected (\textit{e.g.}, $a_{uv}=0$) to another node $v$ in community $j$. Its entries are bounded between $0$ and $1$. Also, recall that in the previous section on the degree regime, the graph sequence is in a constant degree regime if the corresponding matrix $W$ does not scale with $n$, \textit{e.g.}, $s_t=1$.

Here, we look at the behavior of the degree regime through a proposed measure, called FN-differential identifiability, inspired by Amico et al. \cite{amico2018quest}, as follows:
\begin{align}
\mathbb{I}_{diff}^{FNs}&=\mathbb{I}_{self}^{FNs} - \mathbb{I}_{others}^{FNs} \\
&= \langle W_{ii} \rangle - \langle W_{ij} \rangle
\end{align}
where $i,j\in [k]$ and $k=7$ in our study. Moreover, $\langle W_{ii} \rangle$ and $\langle W_{ij} \rangle$ are the averages of the diagonal and off-diagonal entries of matrix $W$, respectively. We formally define $\langle W_{ii} \rangle$ and $\langle W_{ij} \rangle$ to be the differential identifiability within (\textit{e.g.}, $\mathbb{I}_{self}^{FNs}$) and between (\textit{e.g.}, $\mathbb{I}_{others}^{FNs}$) FNs. 

Per Figure \ref{matrixIdiff}, for most threshold values (with the exception of $\t=1$), there is an intensity shift in $W$ from between-FN connectivity to within-FN connectivity strength as $\mathbb{I}_{diff}^{FNs}$ increases across Schaefer parcellation granularity levels. In other words, within-FN identifiability $\mathbb{I}_{self}^{FNs}$ increases as between-FN identifiability $\mathbb{I}_{others}^{FNs}$ decreases. Moreover, the monotonic increase of $\mathbb{I}_{diff}$ also suggests that finer-grain Schaefer parcellations might reflect a higher level of information-theoretic prominence of the \textit{a priori} set of FNs, given the group-average FCs. These results indicate that the Schaefer brain graph sequence resides in the diverging degree regime, with the exception of the trivial case $\t=1$. Empirically, these results suggest that the Schaefer graph sequence is at least not in the constant degree regime, \textit{i.e.}, $s_t \neq 1$.

\section*{Weak Recovery - fMRI Resting State Further Analysis}
\subsection*{Null models}
\begin{figure*}[!ht]
	\centering
	\includegraphics[width=.8\linewidth]{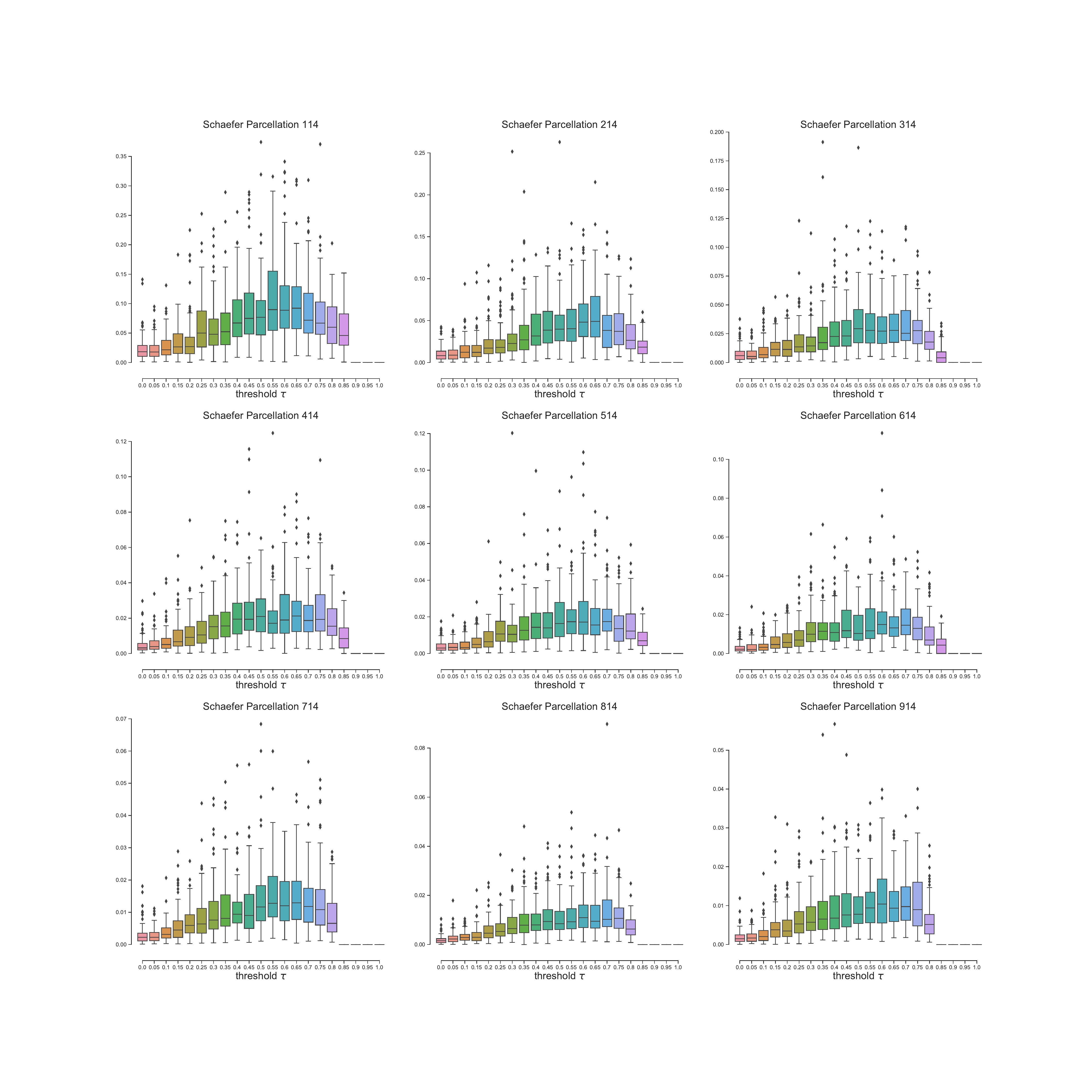}
	\caption{\small Each subplot represents the SNR profiles corresponding to 100 randomized parcellations for each thresholding value $\t \in [0,1]$ for all nine Schaefer parcellations.}
	\label{fig:rand_SNR_dist}
\end{figure*}

The null model is assessed by feeding a randomized partition that respects Yeo's FNs sizes. The number of simulations is 100, and the scanning session is LR. Results on the empirical distribution of randomized SNR scores are shown in Figure \ref{fig:rand_SNR_dist}.

\section*{Methods}
\subsection*{Notations}
In this section, we describe some stochastic block model (SBM) fundamentals, along with some fundamental mathematical notions that are not included in the main text. For instance, a number (scalar) is denoted with a regular letter such as $x,y$; a default vector is denoted by a bold regular letter \textit{i.e.}, $\mathbf{x}$, sometimes with or without a subscripted index, and is in column format. Matrices are denoted by bold, capitalized letters, \textit{i.e.}, $\mathbf{A}$, while a set is denoted by a capitalized letter, \textit{i.e.}, $S$. Further, $\mathbb{N}$ and $\mathbb{R}$ are sets of natural and real numbers, respectively; $[l]$ denotes all positive integers between 1 up to $l$. All other standard mathematical notations are assumed unless otherwise specified.

\subsection*{SBM Definition}
The Stochastic Block Model (SBM) has a history of both depth and breadth, spanning across multiple disciplines. Here, we only extract relevant information regarding SBM literature relevant to our exploration. Stochastic Block Models (SBMs) are random graph models that generate ensembles with clusters. Specifically, they generate Erdos-Renyi (ER) subgraphs union with multi-partite graphs between those subgraphs. A traditional SBM generates binary graphs, \textit{i.e.}, networks with $\cbrac{0,1}$ edges. Nonetheless, there are SBM models developed for weighted networks, which are called weighted SBM (or WSBM).

\subsection*{SBM Inference and Synthesis}
\textbf{(Binary) SBM}. Since SBM is a generative model, it is essential to discuss how to synthesize ensembles using such models, \textit{e.g.}, network synthesis, and how to infer SBM parameters using the observable ensembles, \textit{e.g.}, network inference. In the context of our problem, we have a slightly different starting point as the partition is not latent, though generally, partitions are often inferred. Networks with an existing ground-truth partition are very rare; furthermore, those ground-truths cannot be defined in an absolute sense. The majority of SBMs are defined as follows:
\[  
G \sim SBM(k,p,W, \sigma) 
\]
However, in the context of our paper, the partition is not latent. Specifically, $$(G,\sigma,k)\sim SBM(p,W)$$ for our application.

In the case of $G \sim SBM(k,p,W, \sigma)$, SBM seeks a partition that divides network $G$ into $k$ communities. The probability that two nodes are connected to each other is governed by the probability $W_{\sigma_u,\sigma_v}$. To fit SBM onto a network, one needs to estimate $W=[w_{ij}], \forall i,j\in [k]$ (meaning that $k$ is an \textit{a priori} condition for fitting) along with the community label $\sigma_u,\forall u\in [n]$. Assuming that each edge is drawn independently from identical distributions, the probability that a network $G=A=[a_{uv}]$ is generated (synthesized) from \textit{a priori} $W$ and $\sigma$ (prior beliefs) is as follows:
\[
\P(A \mid W, \sigma) = \prod_{u>v} W^{a_{uv}}_{\sigma_u,\sigma_v} (1-W_{\sigma_u,\sigma_v})^{1-a_{uv}}
\]
for symmetric networks. From the inference standpoint, the Bayesian posterior probability can be computed as follows:
\[
\P(\sigma \mid A) = \frac{\sum_{W} \P(A \mid W, \sigma) \P(W, \sigma)}{\P(A)}
\]
where $\P(W, \sigma)$ represents Bayesian prior beliefs. If there is only one $W$ (\textit{hard constraint}, Piexoto) that is comparable to network $A$ and partition $\sigma$, then we can drop the summation notion, resulting in:
\begin{align*}
\P(\sigma \mid A) &= \frac{\P(A \mid W, \sigma) \P(W, \sigma)}{\P(A)}\\
&= \frac{\exp {\cbrac{-\ln(\P(A \mid W, \sigma)) - \ln(\P(W, \sigma))}}}{\P(A)}
\end{align*}
The hard constraint assumption is a very standard technique to isolate the eventual partition $\sigma$ for inference purposes. Note that the adjacency structure $A$ is, of course, "hard" (there is only one ensemble $A$).

Since $\P(A)$ is also fixed, maximization of posterior probability $\P(\sigma \mid A)$ is equivalent to maximizing 
$$-\ln(\P(A \mid W, \sigma)) - \ln(\P(W, \sigma))$$ 
which is also understood as the minimization of the description length ($DL$, measured \textit{in bits}) of ensemble $A$ using partition $\sigma$. Once again, the hard constraint assumption yields that the description length ultimately only depends on:
$$DL = -\ln(\P(A \mid W, \sigma))$$
In the binary SBM case, it follows that:
\begin{align*}
DL &= -\ln(\prod_{u>v} W^{a_{uv}}_{\sigma_u,\sigma_v} (1-W_{\sigma_u,\sigma_v})^{1-a_{uv}}) \\
&= -\sum_{u>v} a_{uv} \ln(W_{\sigma_u,\sigma_v}) + (1-a_{uv}) \ln(1-W_{\sigma_u,\sigma_v})
\end{align*}
Hence, minimization of $DL$ is equivalent to maximizing the log-likelihood function. 

\textbf{Weighted SBMs}. The assumption of binary edges could be unfitting for some applications, including functional brain networks where there is a need to express different levels of functional coupling strength numerically. In such cases, we need to introduce the structure of covariates (denoted as $x=[x_{\sigma_u,\sigma_v}]$) to model the weights. In this case, the prior is written as follows:
\begin{align*}
\P(x, A \mid \sigma) &= \P(x \mid A, \sigma) \P(A, \sigma)
%	 &= \prod_{u<v} \int \P(x_{\sigma_u,\sigma_v}\mid \gamma) \P(\gamma) d\gamma
\end{align*}
It follows that the posterior probability becomes:
\[
\P(\sigma \mid A, x) = \frac{\P(A \mid x, \sigma) \P(x, \sigma)}{\P(A)}
\]
Using a similar technique in the binary case, one can estimate the covariate structure first so that joint probabilities with $x$, \textit{e.g.}, $\P(x, \sigma)$ and $\P(A, x)$, do not alter the posterior distribution behavior. Hence, the posterior belief is proportional to the priors, which can be written as follows:
\[
\P(\sigma \mid A, x) \sim \P(A \mid x, \sigma)
\]
Ultimately, the task of finding the "ground-truth" partition $\sigma$ depends on the likelihood of prior beliefs. This is equivalent to maximizing $\P(A \mid x, \sigma)$. The first task is, of course, to estimate $x$ as $A$ is already available. We notate the covariate structure $x$ to be more integrated with probability distribution parameter notations $\P(X = x)$. Specifically, we assume that the realized FC edge weights are drawn from some distributions with specific parameter(s).
\begin{figure*}[ht]
	\centering
	\includegraphics[width=\linewidth]{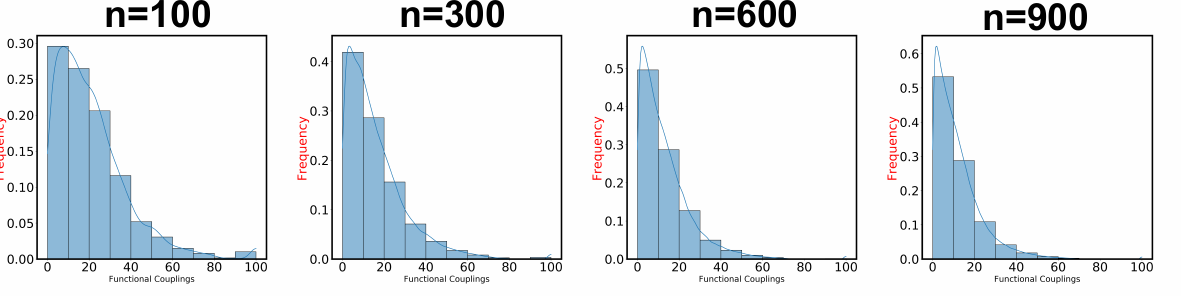}
	\caption{\small Empirical distribution of functional coupling magnitudes (shown as percentage points by multiplying absolute Pearson correlation values by 100) from group-average FCs across four distinct Schaefer parcellations $n=[100,300,600,900]$.}
	\label{fig:GA_FC_histogram}
\end{figure*}

\textbf{Model Selection.} There are different approaches to model selection (\textit{e.g.}, which edge weight distribution one should use, given the empirical data). In the context of this paper, given the empirical distribution of FC edge weight and the usage of absolute functional connectivity (non-negative pair-wise edges), we shortlist two candidate distributions: exponential (continuous) and Poisson (discrete counterpart). Each choice has its pros and cons. For instance, choosing the exponential distribution allows us to stick with continuous ensembles of functional edge weights, which is consistent with how FC edges are computed using Pearson correlations. However, in a continuous distribution, the probability of an FC edge taking on a particular value is zero by definition. Yet, the functional connectome is sparse \cite{betzel2018diversity}, \textit{e.g.}, the majority of pairwise interactions between two brain regions are non-existent. Hence, using the exponential distribution will not suffice. A common approach is to use a different distribution such as the Binomial to model edge (non-)existence and connect the two distributions using a weighted average (see the methods section in \cite{betzel2018diversity} for further details). This will force the modeler to make a precursor assumption on weight value, which is not ideal.

On the other hand, using the Poisson distribution (a discrete counterpart of the exponential distribution) offers us the distinct advantage of modeling a non-zero probability of getting zero-valued functional edges. Recall that the Poisson probability density function is as follows:
\[
f(k, \lambda) = \P(X = k) = \frac{\lambda^k e^{-\lambda}}{k!}
\]
Clearly, $\P(X = 0) = e^{-\lambda} > 0$, which ultimately depends on $\lambda$ inference based on empirical observations of functional edges. Nonetheless, the shortcoming of using a discrete distribution is precisely the advantage of using the exponential one: being able to model edges in a continuous manner. To overcome this shortcoming of discrete distribution usage, we convert functional couplings (computed by Pearson correlations, which are nicely bounded between $[-1, 1]$) to percentage points and round to the nearest integer. For instance, if a functional edge has a value of $a_{uv} = 0.588$, the weighted graph will take $a_{uv} = 59$. The reason for rounding to the nearest integer is that the Poisson distribution takes on non-negative integer values $\mathbb{N}^{+}$. Note that using the Poisson Distribution makes no essential topological changes to the original FC other than rounding functional couplings to the nearest integer.

In this paper, we use the Poisson distribution for degree sequence as proposed by Karrer and Newman \cite{karrer2011stochastic}. In the next sections, we review the inference procedure (as proposed in \cite{karrer2011stochastic}) for both assumptions:
\begin{itemize}
    \item Non-degree-corrected WSBM;
    \item Degree-Corrected WSBM.
\end{itemize}
\textbf{WSBM Inference Procedure:} In this paper, we use the method described at \url{https://graph-tool.skewed.de/} by Tiago Peixoto. Further treatments on weighted SBM can be found in \cite{peixoto2018nonparametric}. After reviewing the inference approaches, we compare the philosophical similarities and differences between WSBM inference and $Q$ score modularity.

\subsubsection*{Standard (Non-degree-corrected) WSBM}
The review of non-degree-corrected (NDC) WSBM is provided in a well-cited paper by Karrer and Newman \cite{karrer2011stochastic}. The authors assumed such a distribution for multi-graph ensembles, where edges can take on integer values larger than 1. In this case, the prior probability can be written as follows:
\begin{align*}
\P(A\mid x,\sigma) &= \prod_{u<v} \frac{(x_{\sigma_u,\sigma_v})^{A_{uv}} e^{-x_{\sigma_u,\sigma_v}}}{A_{uv}!}\\
& \times \prod_{u} \frac{(x_{\sigma_u,\sigma_u})^{A_{uu}/2} e^{-x_{\sigma_u,\sigma_u}}}{(A_{uu}/2)!}
\end{align*}
It is important to note that the expected adjacency structure in this case is 
$$\E(A_{NDC})= YxY^\top$$ 
where $Y \in [0,1]^{n\times k}$ is the node community membership matrix, \textit{i.e.}, $y_{ul}=1$ if and only if node $u$ is in community $l \in [k]$.

 Note that self-loop edge weight cannot be counted twice. For symmetric networks where $A_{uv}=A_{vu}$ and $x_{ij}=x_{ji}$, the above prior probability can be written as follows:
\begin{align*}
\P (A\mid x,\sigma) &= \frac{\prod_{ij} x_{ij}^{C_{ij}/2} \exp(-\frac{1}{2} |\Omega_i||\Omega_j| x_{ij})}{\prod_{u<v} (A_{uv}! \prod_{u} 2^{A_{uu}/2} (A_{uu}/2)!)}
\end{align*}
where $|\Omega_i|$ is the cardinality of community $i$, $C_{ij}$ is the counted number of edges between community $i$ and $j$ which can be simply computed by:
\[
C_{ij} = \sum_{u,v} A_{uv} \delta_{\sigma_u,i} \delta_{\sigma_v,j}
\]
where $\delta$ is the Kronecker delta function as defined in the main text. Similar to the binary case, the log-function is then be:
\[
\log \P(A\mid x,\sigma) = \sum_{ij} (C_{ij} \log(x_{ij})-|\Omega_i||\Omega_j|x_{ij}) + \Theta(G)
\]
where $\Theta(G)$ is the quantity dependent on ensemble $G$ (such as $|\Omega_i|$ or $A_{uv}$) which has no impact onto the logarithmic function behavior (\textit{i.e.} not impacting the optimal value of this function). The inference process reduces to maximizing:
\begin{align*}
L(x,\sigma) &= \sum_{ij} (C_{ij} \log(x_{ij})-|\Omega_i||\Omega_j|x_{ij}) 
%	 					& =  \sum_{ij} (C_{ij} log(x_{ij})) - 2m 
\end{align*}
Note that here, we drop $A$ (ensemble adjacency structure) just to ease notation usage and emphasize which variable(s) the likelihood function depends on. To optimize the above function, one can use differential calculus as follows:
\[
\frac{dL}{dx_{ij}}=L_{ij}^{'} = \frac{d}{dx_{ij}} \left[ \frac{C_{ij}}{x_{ij}} - |\Omega_i||\Omega_j| \right]
\]
Setting the first derivative to zero, \textit{e.g.} $L'=0$, we obtain:
\[
\hat{x}_{ij}= \frac{C_{ij}}{|\Omega_i||\Omega_j|}
\]
Note that now we have estimated $x$, \textit{i.e.}, the covariate structure, the likelihood function can be written as follows:
\begin{align*}
L(\hat{x},\sigma) = \sum_{ij} (C_{ij} \log(x_{ij})) - 2m
\end{align*}
Dropping the constant $2m$ (ensemble node's degree sum) and substituting the estimated covariate $\hat{x}$, the log-likelihood function can now be written as:
\begin{align*}
L(\sigma) &= \sum_{ij} C_{ij} \log \left( \frac{C_{ij}}{|\Omega_i||\Omega_j|} \right)
\end{align*}
Using simple algebra, the log-likelihood function can be rewritten as:
\begin{align*}
L(\sigma) &= 2m \sum_{ij} \frac{C_{ij}}{2m} \left[ \log \left( \frac{C_{ij}/2m}{|\Omega_i||\Omega_j|/n^2} \right) - \log \left( \frac{n^2}{2m} \right) \right]\\
&= \sum_{ij} \frac{C_{ij}}{2m} \log \left( \frac{C_{ij}/2m}{|\Omega_i||\Omega_j|/n^2} \right) + \Theta(G)
\end{align*}
where, again, $\Theta$ is a constant function based on ensemble $G$.

Let $Y$ and $Z$ be the random variables representing community assignment on one end of a stub (half-edge). Then we can build a joint probability distribution between $Y'$ and $Z'$ as follows:
$$\P_\sigma=\P_\sigma (Y',Z')=\frac{C_{ij}}{2m},\quad \forall i,j\in [k]$$ 	 
On the other hand, the randomized counterpart distribution of these random variables (with the same \textit{a priori} partition $\sigma$) is 
$$\P^{WSBM}_{\text{null}} = \frac{|\Omega_i||\Omega_j|}{n^2}$$
In this case, edge formation (from two stubs) is completely random with probability $\frac{|\Omega_i|}{n}$ and $\frac{|\Omega_j|}{n}$ for each stub. Overall, the likelihood function becomes:
\begin{align*}
L(\sigma) &= \sum_{ij} \P_\sigma (ij) \log \left\lbrace \frac{\P_{\sigma} (ij)}{\P^{WSBM}_{\text{null}} (ij)} \right\rbrace\\
&= \sum_{ij} \P_\sigma (ij) \left\lbrace \log( \P_{\sigma} (ij)) - \log( \P^{WSBM}_{\text{null}} (ij)) \right\rbrace
\end{align*}

On the other hand, the Kullback-Leibler Divergence between two probability distributions $P(x)$ and $Q(x)$ is defined to be:
\[
D_{KL} (P\mid\mid Q) = \sum_{x\in \mathcal{X}} P(x) \log \left\lbrace \frac{P(x)}{Q(x)} \right\rbrace
\]
where $x\in \mathcal{X}$ is the random variable taking on values in the sample space $\mathcal{X}$. Then the log-likelihood function above can be thought of as an information theoretic measurement between the "ground-truth" probability distribution $x(\sigma)$ and the corresponding null distribution $x(\text{null})$. If we only look at what happens within communities, the quality function becomes:
\[
L_{\text{within}}(\sigma) =  \sum_{i} \P_\sigma (ii) \left\lbrace \log( \P_{\sigma} (ii)) - \log( \P^{WSBM}_{\text{null}} (ii)) \right\rbrace
\] 
If we substitute the estimated $\P_\sigma$ and $\P_{\text{null}}$ above into this equation, we obtain:
\[
L_{\text{within}}(\sigma) = \sum_{i} \frac{C_{ii}}{2m} \left\lbrace \log \left( \frac{C_{ii}}{2m} \right) - \log \left( \frac{|\Omega_i|^2}{n^2} \right) \right\rbrace
\]
Of course, we also have the log-function describing the differential information description requirement between communities:
\[
L_{\text{between}}(\sigma) = \sum_{i\neq j} \frac{C_{ij}}{2m} \left\lbrace \log \left( \frac{C_{ii}}{2m} \right) - \log \left( \frac{|\Omega_i||\Omega_j|}{n^2} \right) \right\rbrace
\]
We will compare the within-community $L_{\text{within}}$ with the modularity function $Q$ score in the subsequent sections.

\subsubsection*{Degree-corrected WSBM}
For the degree-corrected (DC) WSBM case, a new hyper-parameter is introduced into the model $\theta_r$ (arbitrary constant terms that are $o(x_{\sigma_r,\sigma_s})$, i.e., constant terms that get absorbed into $x_{ij}$). The prior probability can now be written as follows:
\begin{align*}
\P(A\mid \theta,x,\sigma) &= \prod_{u<v} \frac{(\theta_u \theta_v x_{\sigma_u,\sigma_v})^{A_{uv}} \exp(-\theta_u \theta_v x_{\sigma_u,\sigma_v})}{A_{uv}!}\\	
&\times \prod_u  \frac{(\theta_u^2 x_{\sigma_u,\sigma_u})^{A_{uu}/2} \exp(-\theta_u^2 x_{\sigma_u,\sigma_u})}{(A_{uu}/2)!}
\end{align*} 
where $\sum_{u} \theta_u \delta_{\sigma_u,i}=1$ (with $\delta$ as the Kronecker delta function as usual). Basically, $\theta_u$ represents the probability that a half-edge (stub) in community $i$ originated from $u$ itself, where $\sigma_u=i$. It is noteworthy that the expected value of the adjacency structure in this case is no longer just $x_{\sigma_u,\sigma_v}$ but instead:
\begin{align*}
\E (A_{DC}) &= [\E(a_{uv})] = \theta_u x_{\sigma_u,\sigma_v} \theta_v\\
&= \diag(\theta) YxY^T \diag(\theta)
\end{align*}
where $\diag(\theta)=\diag([\theta_u])$ is the diagonal matrix containing the $\theta_u$ weights of node $u$. The priors can then be condensed as follows:
\[
\P (A\mid \theta,x,\sigma) = \frac{\prod_u \theta_u^{d_u} \prod_{ij} x_{ij}^{C_{ij}/2} \exp(-\frac{1}{2}x_{ij})}{\prod_{u<v} A_{uv}!\prod_u 2^{A_{uu}/2} (A_{uu}/2)!}
\]
with $d_u$ being the degree of node $u$.
The log-likelihood function is then
\begin{align*}
L &= \log \P (A\mid \theta,x,\sigma)\\
&= 2\sum_u d_u \log\theta_u + \sum_{ij} \cbrac{C_{ij} \log x_{ij} - x_{ij}}\\
&= L_1 + L_2
\end{align*}
where $d_u$ is the degree of node $u$, and again, constant terms $\Theta(G)$, which contain $A_{uv}$ terms, are ignored. The goal is to maximize this log-function, compartmentally, with respect to the normalization condition $\sum_{u} \theta_u \delta_{\sigma_u,i}=1$. We look at them separately (again, ignoring any constants). Maximizing $L_2 = \sum_{ij} \cbrac{C_{ij} \log x_{ij} - x_{ij}}$ is straightforward by taking the derivative with respect to $x_{ij}$. Specifically,
\[
L'_2 = \frac{dL_2}{dx_{ij}} = \frac{C_{ij}}{x_{ij}} - 1 = 0 \rightarrow \hat{x}_{ij} = C_{ij}
\]
\begin{align*}
L_1 &= \sum_u d_u \log\theta_u = \sum_i d_u \delta_{\sigma_u,i} \log\theta_u\\
&= \sum_{i} \cbrac{\sum_{u\mid \sigma_u=i} d_u \log\theta_u}\quad s.t.\quad \sum_{u\mid \sigma_u=i} \theta_u = 1\\
&= \sum_{i} \cbrac{ s_i \sum_{u\mid \sigma_u=i} \frac{d_u}{s_i} \log\theta_u} \quad s.t.\quad \sum_{u\mid \sigma_u=i} \theta_u = 1\\
\end{align*}
where $s_i = \sum_{u\mid \sigma_u=i} d_u$ is the number of half-edges in community $i$. Note that there are $|\Omega_i|$ terms of $\theta_u$ for each community. We see that $L_1$ is the entropy of the probability distribution representing the random variable $\theta$, i.e., the probability that an edge in community $i$ lands on $u$ for which $\sigma_u=i,\forall i$. This entropy is minimized when  
\[\hat{\theta}_u=\frac{d_u}{\sum_u d_u}\]
Here, it is important to note that if we choose a random uniform distribution for the random variable $\theta$ (e.g., $\hat{\theta}_u=\frac{1}{|\Omega_i|}$), we obtain minimized $L_1$, which reduces $L$.

\subsection*{Difference between Non-degree-corrected and Degree-corrected Models}
Plugging in the estimated parameters for both cases of WSBM, we obtain:
\[
L_{NDC} = \sum_{ij} C_{ij} \log\sbrac{\frac{C_{ij}}{|\Omega_i||\Omega_j|}}
\]
and 
\begin{align*}
L_{DC} &= \sum_{ij} C_{ij} \log\sbrac{\frac{C_{ij}}{s_i s_j}}\\ 
&= \sum_{ij} \frac{C_{ij}}{2m} \log \sbrac{\frac{C_{ij}/2m}{(s_i/2m)(s_j/2m)}}
\end{align*}
which is the Kullback-Leibler divergence between $P(x)$ (same as in the NDC case) and $Q_{DC} (x)$. In other words,
\[
\P^{WSBM}_{null} = \frac{s_i s_j}{(2m)^2}
\] 
for the DC case. Recall that for the NDC case, the null model is:
\[
\P^{WSBM}_{null} = \frac{|\Omega_i||\Omega_j|}{n^2}
\]
Thus, the best fit to the NDC WSBM is the partition that most surprises the Erdos-Reyni random counterpart while for the DC WSBM case, it is the group assignment that is most surprising to the random model with the same empirical degree sequence.

\subsection*{Modularity}
The $Q$-score for a given partition can be computed as follows:
\begin{align*}
Q(\sigma,\alpha=1) = Q(\sigma) &= \sum_{u,v} (A_{uv} -\a P_{uv})\delta(\sigma_u,\sigma_v)\\
&= \frac{1}{2m} \sum_{uv} \cbrac{A_{uv}-\frac{d_u d_v}{2m}} \delta(\sigma_u, \sigma_v)
\end{align*}
where $2m$ and $d_u$ represent the graph and node degrees, respectively:
$$\forall u\in V(G): d_u=\sum_{v} A_{uv}\quad \& \quad 2m = \sum_{u} d_u$$
and the default scaling factor $\a=1$; this scaling factor is typically used to scan the hierarchical structure of communities in a network.

There is more than one way to model the null model \( P_{uv} \). Newman's approach is \( P_{uv} = \frac{d_u d_v}{2m} \), which represents the random graph (no particular community structures) with the same empirical degree sequence. In theory, one can assume Poisson Distribution for node degree (like in the WSBM case). In the case of Newman's \( Q \), this null model is built with respect to the empirical network degree. Furthermore, the tuning parameter is, by default, set at \( \alpha = 1 \), and the delta function is defined as
\[ \delta(\sigma_u, \sigma_v) = \begin{cases}
1, & \text{if } \sigma_u = \sigma_v \\
0, & \text{if } \sigma_u \neq \sigma_v
\end{cases} \]

If \textit{a priori} partition \( \sigma \) is known, then the modularity score can be written in a blockage format (only surviving terms in within communities: \( \forall u,v \in V(G) \mid \sigma_u = \sigma_v = i \in [k] \)) as follows:
\begin{align*}
Q(\sigma) &=  \frac{1}{2m} \sum_{uv} \cbrac{A_{uv} - \frac{d_u d_v}{2m}} \delta(\sigma_u, \sigma_v)\\
&= \sum_{i=1}^{i=k} \sbrac{\frac{\sum_{u,v\in i}A_{uv}}{2m} - \frac{\sum_{u,v\in i}d_u d_v}{(2m)^2} }\\
&=  \sum_{i=1}^{i=k} \sbrac{\frac{C_{ii}}{2m} - \sum_{u,v\in i} \frac{d_u}{2m} \frac{d_v}{2m} }\\
&= \sum_{i=1}^{k} \sbrac{\frac{C_{ii}}{2m} - \sbrac{\frac{s_i}{2m}}^2}\\
%	 &= \sum_{i=1}^{k} \sbrac{\frac{C_{ii}}{2m} - \cbrac{\frac{|\Omega_i|^2}{n^2}}^2}\\
&= \sum_{i=1}^{k} (\P_\sigma (ii) - \P^{Q}_{null} (ii))
\end{align*}
Because:
\[
\sum_{u,v\in i}A_{uv} = C_{ii}
\] 
and 
\[
\sum_{u,v\in i}d_u d_v = \sum_{\sigma_u=i} d_u^2 + 2\sum_{u\neq v} d_u d_v = (\sum_{\sigma_u=i} d_u)^2 = (s_1)^2
\]
where \( s_1 \) is the total number of half-edges (stubs) that originate from nodes in community \( i \).

We also look at the null model from another perspective: the event of a stub (half-edge) exists at node \( u \) with probability \( \P_u(\text{stub}) = \frac{d_u}{2m} \), likewise, at node \( v \) with \( \P_v(\text{stub}) = \frac{d_u}{2m} \). These two independent events need to happen sequentially to form an edge between node \( u \) and \( v \) with probability 
\[ \P^{Q}_{\text{null}} = \P_{uv}(\text{edge}) = \frac{d_u}{2m} \frac{d_v}{2m}, \forall \sigma_u = \sigma_v = i \in [k] \]
Note that here, no community indication is available for either node \( u \) or \( v \), which implies a null model (i.e., random partition) of \( \sigma \) (ground-truth).

The \( Q \)-score can be applied to both binarized or weighted graphs. In this case, for each threshold value, the \( Q \)-score is computed for the weighted group-average FCs across Schaefer granularity levels. Maximizing modularity has been shown to unravel assortative communities, while SBM has been shown to uncover different types of communities, beyond assortative ones \cite{betzel2018diversity}. 

\subsection*{Philosophical Similarity between Log-likelihood Function and \( Q \)-score}
In this section, we compare the NDC, DC WSBM, and \( Q \)-score approach by first revisiting their formulas:
\begin{enumerate}
    \item The NDC log-likelihood function (within communities): \[ L_{\text{within}} = \sum_{i} \frac{C_{ii}}{2m} \cbrac{\log \sbrac{\frac{C_{ii}}{2m}} - \log\sbrac{\frac{|\Omega_i|^2}{n^2}}} \]
    \item The DC log-likelihood function (within communities): \[ L_{\text{within}} = \sum_{i=1}^{k} \frac{C_{ii}}{2m} \cbrac{\log (C_{ii}/2m) - \log\sbrac{\frac{s^2_i}{(2m)^2}}} \]
    \item \( Q \)-score modularity function, using community block format: \[ Q = \sum_{i=1}^{k} \sbrac{\frac{C_{ii}}{2m} - \sbrac{\frac{s_1}{2m}}^2} \]
\end{enumerate}

It is very interesting (yet, not surprising) that the two most well-known methods for community detection are based on a similar principle of comparing a structure-less counterpart (that has some similar topological characteristics) with the network at hand (which is hypothesized to have some latent structure of communities). In both approaches, the hypothesized distribution of random variables \( Y' \) and \( Z' \) for within-community is actually the same:
\[ \P^{\text{NDC-WSBM}} = \P^{\text{DC-WSBM}} = \P^{Q} = \frac{C_{ii}}{2m} \]
Obviously, there is a difference between the null model choice between the \( Q \) score and NDC WSBM approach as the latter does not feature the observed network degree sequence while Newman's \( Q \) method actually does. This shortcoming is resolved with the DC WSBM approach as mentioned in the previous section. In fact, the DC WSBM and \( Q \) score null model is actually the same. Specifically, for DC WSBM, the null model is
\[ \P^{\text{DC-WSBM}}_{\text{null}} = \frac{s_i^2}{(2m)^2} \]
while for the modularity approach, it is also:
\[ \P^{Q}_{\text{null}} =  \frac{s_i^2}{(2m)^2} \]
What, then, is the shortcoming of the \( Q \) score approach? It does not emphasize what happens with the "between" community dynamics. To be clear, one can rewrite the \( Q \) score so that it reflects between-community edges as proposed by Fortunato \cite{fortunato2010community} as follows:
\begin{align*}
Q &= \sum_{i=1}^{k} \sbrac{\frac{C_{ii}}{2m} - \sbrac{\frac{s_1}{2m}}^2}\\
&= \frac{-1}{m}\sbrac{\cbrac{m-\frac{1}{2}\sum_i C_{ii}} - \cbrac{m - \sum_i (\frac{s_i^2}{4m})}}\\
&= \frac{-1}{m} \sbrac{Cut - \E (Cut)}
\end{align*}
where \( Cut = \cbrac{m-\frac{1}{2}\sum_i C_{ii}} \) being the number of inter-community edges and \( \E(Cut) \) is its corresponding expected counterpart. Basically, modularity would like to maximize the within-community edges (equivalently, minimize the between-community edges). Hence, it biases towards assortative community assignments.

On the other hand, the log function of WSBM has also incorporated what happens between communities through \( L_{between} \). This is why SBM inference method shines over traditional \( Q \) maximization techniques for its ability to uncover a more diverse range of community structures beyond assortative ones.

\subsection*{Statistics to compare \( Q_{\text{max}} \) and SBM-inference community detection approaches}
Given the node set of \( N \) elements \( S = \{ s_i \mid i \in [N] \} \), to quantify the amount of shared information between two partitions (e.g., \( \sigma^1 = \{ \sigma_r^1 \mid r \in [R] \} \) and \( \sigma^2 = \{ \sigma_c^2 \mid c \in [C] \} \)), a common approach would be using mutual information score, which can be quantified as follows:
\[ MI(\sigma^1,\sigma^2) = \sum_{r=1}^{R} \sum_{c=1}^{C} \P_{\sigma^1\sigma^2} (r,c) \log \frac{\P_{\sigma^1\sigma^2} (r,c)}{\P_{\sigma^1}(r) \P_{\sigma^2} (c)} \]
where \( R \) and \( C \) are the number of clusters in partition vectors \( \sigma^1 \) and \( \sigma^2 \), respectively; \( \P_{\sigma^1\sigma^2} \) and \( \P_{\sigma^1} \) are the joint and marginal probability distributions, respectively, between two discrete random variables representing two realized partitions. A typical initialization step is to build a contingency table which indicates the number of common nodes has in common between cluster \( \sigma^1_r \) and \( \sigma^2_c \):
\[ \begin{pmatrix} n_{11} & n_{12} & \cdots & n_{1C}\\ \vdots & \cdots & \vdots & \cdots\\ n_{R1} & n_{R2} & \cdots & n_{RC} \end{pmatrix} \]
where \( n_{rc} \) represents the number of common entities between cluster \( \sigma^1_r \) and \( \sigma^2_c \); the row and column marginal sums are denoted as \( \vec{a} = a_r \) and \( \vec{b} = b_c \), respectively. By construction, \( \sum_r a_r = \sum_c b_c = N \). The expected mutual information for a random partition with the same contingency table has a closed-form formula as proposed in \cite{vinh2010information}:
\begin{align*}
\E (MI(\sigma^1,\sigma^2)) &= \sum_{r} \sum_{c} \sum_{\max(1,a_r+b_c-N)}^{\min(a_r,b_c)} \frac{n_{rc}}{N} \log\cbrac{\frac{N \times n_{rc}}{a_r b_c}}\\
& \times \frac{a_r! b_c! (N-a_r)! (N-b_c)!}{N!n_{rc}! (a_r-n_{rc})!(b_c-n_{rc})! (N-a_r-b_c+n_{rc})!}
\end{align*} 
The entropy associated with the two partitions are: 
\[ H(\sigma^1) = \sum_{r=1}^{R} \P_{\sigma^1} (r) \log(\P_{\sigma^1} (r)) \]
where \( \P_{\sigma^1} (r)= \frac{|\sigma^1_{r}|}{N} \) is the probability that an element picked randomly from set \( S \) belongs to \( \sigma^1_r \). Analogously,
\[ H(\sigma^2) = \sum_{j=1}^{C} \P_{\sigma^2} (j) \log(\P_{\sigma^2} (j)) \]
Finally, putting all components together, adjusted (for chance) normalized mutual information (AMI) can be computed as follows:
\[ AMI = \frac{MI-\E(MI)}{\max(H(\sigma^1),H(\sigma^2))-\E(MI)} \]
Note that there are other ways to average the independent entropy of the two partitions such as arithmetic. In this paper, we use the maximum between the two entropy quantities.
\subsection*{Graph Sequence and Required Topological Features for Recovery}
A graph sequence is a mathematical series of graph ensembles generated by some fixed rules, denoted by $\{G_{l}\}$ where $l$ is the sequence index. For example, one can generate an ER random graph sequence denoted as $G_t (t,p)$ with fixed $p$ and its limiting graph denoted as $G_{t=\infty}(t,p)$, also known as a graphon. In recent developments in $SBM(k,p,W)$ theory, it is important to note the theoretical scaling characteristics of model parameters such as $k$, $p$, and $W$, and make necessary assumptions, i.e., which scales with $t$, and which stays constant as the graph size grows to $\infty$.

Firstly, it is common to assume that $p$ and $k$ do not scale with $t$; hence, the number of communities and their respective sizes do not grow with $t$ \cite{abbe2017community}. In other words, communities are assumed to have linear sizes \cite{abbe2017community}. Moreover, matrix $W$ has theoretical ties with an important topological characteristic of a graph sequence, such as the degree regime.

The importance of the degree regime lies in its relations with graph connectivity. There are two important degree regimes relevant for graph partition recoverability:
\begin{itemize}
    \item \textbf{Constant Degree Regime}: In this regime, the connectivity pattern is fixed (independent of Schaefer granularity levels). Asymptotically, node degrees do not scale with graph size, i.e., $W=O(n^{-1})$. In random graph theory, this is the degree where an ER graph is expected to have a giant component. This regime satisfies the minimum requirement for the weak-recovery criteria, which will be formally defined later.
    \item \textbf{Diverging Degree Regime}: In this regime, the connectivity pattern varies with graph sequence sizes. Asymptotically, node degrees scale with graph size at a scalable factor $s_t$, i.e., $W=O(\log(n) n^{-1})$. In random graph theory, this degree regime generates a connected ER ensemble, in expectation. This regime satisfies the minimum requirement for exact recovery, which will be defined in a later section.
\end{itemize}

\section*{Data, Atlases, and Code Availability}

\subsection*{Neuroimaging Data Acquisitions}
The fMRI dataset used in this paper is available in the Human Connectome Project (HCP) repository (\url{http://www.humanconnectome.org/}), Released Q3. The processed functional connectomes obtained from this data and used for the current study are available from the corresponding author upon reasonable request. Please refer to the detailed descriptions below on the dataset and data processing.

We first describe the acquisitions of raw neuroimaging data from 409 Unrelated Subjects chosen from the list of 1200 participants by Essen et al. \cite{van2012human,van2013wu} in the Human Connectome Project (HCP) release. This subset of participants ensures that no two participants have any family relations, sharing parents or being siblings. This selection is particularly critical to avoid any confounding effects in our subsequent analyses, such as group average analysis, due to family structures.   

Per HCP protocol, all subjects gave written informed consent to the HCP consortium. The two resting-state functional MRI acquisitions (HCP filenames: rfMRI\_REST$_1$ and rfMRI\_REST$_2$ were acquired in separate sessions on two different days, with two distinct scanning patterns (left to right and right to left) in each day, \cite{glasser2013minimal}, \cite{van2012human}, and \cite{van2013wu} for details. This release also includes data from seven different fMRI tasks: gambling (tfMRI\_GAMBLING), relational reasoning (tfMRI\_RELATIONAL), social (tfMRI\_SOCIAL), working memory (tfMRI\_WM), motor (tfMRI\_MOTOR), language (tfMRI\_LANGUAGE, including both a story-listening and arithmetic task), and emotion (tfMRI\_EMOTION). Per \cite{glasser2013minimal}, \cite{barch2013function}, three tasks MRIs are obtained: working memory, motor, and gambling. The local Institutional Review Board at Washington University in St. Louis approved all the protocols used during the data acquisition process. Please refer to \cite{glasser2013minimal,barch2013function,smith2013resting} for further details on the HCP dataset.

\subsection*{Constructing Functional Connectomes}
We used the standard HCP functional preprocessing pipeline, which includes artifact removal, motion correction, and registration to standard space, as described in \cite{glasser2013minimal,smith2013resting} for this dataset. For the resting-state fMRI data, we also added the following steps: global gray matter signal regression; a bandpass first-order Butterworth filter in both directions; z-scores of voxel time courses with outlier eliminations beyond three standard deviations from the first moment \cite{marcus2011informatics,power2014methods}. 

For task fMRI data, the aforementioned steps are applied, with a relaxation for the bandpass filter [0.001 Hz, 0.25 Hz]. Starting from each pair of nodal time courses, Pearson correlation is used to fill out the functional connectomes for all subjects at rest and seven designated tasks. This would yield symmetrical connectivity matrices for all fMRI sessions.

\subsection*{Brain Atlases}
The brain atlases used in this work are sequential, in the sense that their granularity increases, ranging from 100 nodes to 900 nodes (increment of 100 nodes each time), registered on the cortical surface of the brain. These sequential atlases are made possible thanks to the work of Schaefer and colleagues \cite{schaefer2018local}. Similarly to references \cite{amico2018quest,amico2018mapping}, 14 sub-cortical regions were added, as provided by the HCP release (filename $Atlas$\_$ROI2.nii.gz$). We accomplished this by converting this file from NIFTI to CIFTI format using the HCP Workbench software [\url{(http://www.humanconnectome.org/software/connectome workbench.html)}, with the command \texttt{-cifti- create-label}. The resultant sizes of ROI-based connectomes are, hence, 114, 214,..., 914 nodes for rest and any given fMRI tasks. Mathematically, we denote the Schaefer parcellation sequence to be $G_{t_\ell}$ where $\ell\in [9]$ and $t_\ell=[114,214,...,914]$.

Moreover, Schaefer parcellations are also coupled nicely with further subdivisions of Yeo's functional networks \cite{yeo2011organization} so that the partition associated with a coarser Schaefer graph is related to that of a finer-grained Schaefer one. For a fixed Schaefer granularity (indexed $t_\ell$), we denote the corresponding Yeo's resting-state networks to be $\sigma_{t_\ell}$.

For instance, let $u_{114}$ be a node in the Schaefer graph with $n=114$ nodes and a community label 
$$\sigma_{114}=\{u_{114}\mapsto i\mid i\in [k], \forall u_{114}\in [114]\}.$$ 

Say, this is further subdivided into two nodes $v^{'}_{214}$ and $v^{''}_{214}$ in the next Schaefer graph in the sequence, i.e., $G$ with $n=214$ nodes, and that $u_{114}=\{v^{'}_{214},v^{''}_{214}\}$. Then, it follows that: 
\[
\sigma_{214}=\{v^{'}_{214}\mapsto i \quad \& \quad v^{''}_{214}\mapsto i\mid i\in [k],\forall v_{214}\in [214]\}
\]
In fact, we can generalize this as follows: 
\begin{definition}
    Let $l,q$ be graph sequence indices and $\sigma$ be the network partition. If 
    \begin{enumerate}
        \item $u_{n_l}=\cup u_{n_q} \quad \text{s.t.}\quad l<q,\quad u_{n_l}\in V(G_{n_l}),\quad u_{n_q}\in V(G_{n_q})$;
        \item $\sigma_{n_l}=\{u_{n_l} \mapsto i\mid i \in [k]\}$
    \end{enumerate}
    Then, $\sigma_{n_q}=\{u_{n_q} \mapsto i\mid i \in [k]\}$.
\end{definition}
In practice, the subsequent divisions from coarser to finer granularity of Schaefer parcellations are not perfectly hierarchical in the sense that one node in the coarser parcellation does not perfectly parcellate into subsequently smaller ROIs in the finer one. Nonetheless, in this context, we can relax the condition (i) as follows: if the node associated with the coarser parcellation has the majority of spatial overlaps with the ones in subsequently finer parcellations of the Schaefer graph sequence, then they are assigned to the same resting state network.
	
% Next, in the upcoming sections, we introduce the definition of recovery criteria with respect to ground-truth (or highly putative) partition in network. We then move on to present and interpret the corresponding theorems. Although there are more then two types of recovery ( ground-truth partition), we focus on two types of partitions: weak and exact recovery. 	

\subsection*{Code Availability}
The code to perform the \textit{reconFC} procedure can be found at\\ 
\href{https://github.com/ngcaonghi/fc_threshold_framework}{\color{blue} https://github.com/ngcaonghi/fc\_threshold\_framework}. 

\newpage
\Large{\textbf{References}}\\
\bibliographystyle{abbrv}
\bibliography{SBM}
%\eject